\documentclass[a4paper,12pt, epsfig]{article}
\usepackage{epsfig}
\usepackage{amssymb}
\usepackage{amsfonts}
\usepackage{amsmath}
\usepackage{graphicx}
\newskip\humongous \humongous=0pt plus 1000pt minus 1000pt

\newif\ifdtup

\catcode`\@=11

\@addtoreset{equation}{section}
\def\theequation{\thesection.\arabic{equation}}

\def\@normalsize{\@setsize\normalsize{15pt}\xiipt\@xiipt
\abovedisplayskip 14pt plus3pt minus3pt%
\belowdisplayskip \abovedisplayskip
\abovedisplayshortskip \z@ plus3pt%
\belowdisplayshortskip 7pt plus3.5pt minus0pt}

\def\small{\@setsize\small{13.6pt}\xipt\@xipt
\abovedisplayskip 13pt plus3pt minus3pt%
\belowdisplayskip \abovedisplayskip
\abovedisplayshortskip \z@ plus3pt%
\belowdisplayshortskip 7pt plus3.5pt minus0pt
\def\@listi{\parsep 4.5pt plus 2pt minus 1pt
     \itemsep \parsep
     \topsep 9pt plus 3pt minus 3pt}}

\relax

\catcode`@=12

\catcode`\@=11

\def\section{\@startsection{section}{1}{\z@}{3.5ex plus 1ex minus
   .2ex}{2.3ex plus .2ex}{\large\bf}}

\def\thesection{\arabic{section}}
\def\thesubsection{\arabic{section}.\arabic{subsection}}

\def\appendix{\setcounter{section}{0}
 \def\thesection{Appendix \Alph{section}}
 \def\thesubsection{\Alph{section}.\arabic{subsection}}
 \def\theequation{\Alph{section}.\arabic{equation}}}
\def\SymBoxes#1#2#3#4{\newdimen\un@t \un@t#3%
\raisebox{#1}{\rule{#2\un@t}{#4}\hskip-#2\un@t% lower horizontal
\@tempdimb\un@t \advance\@tempdimb by-#4\@tempcntb#2\relax%
\@whilenum{\@tempcntb>0}\do{%                         % #2 vertical lines
\rule{#4}{\un@t}\hskip\@tempdimb \advance\@tempcntb by\m@ne}%
\hskip-#2\un@t \rule[\un@t]{#2\un@t}{#4}%
\rule[\un@t]{#4}{#4}\hskip-#4%             % upper horizontal line
\rule{#4}{\un@t}}\hskip-#4}                % rightest vertical line
\begin{document}
%\begin{letter}{~}

%%%%%%Define some new commands and  macros
\newcommand{\beq}{\begin{equation}}
\newcommand{\eeq}{\end{equation}}
\newcommand{\bea}{\begin{eqnarray}}
\newcommand{\eea}{\end{eqnarray}}
\newcommand{\beas}{\begin{eqnarray*}}
\newcommand{\eeas}{\end{eqnarray*}}
\newcommand{\defi}{\stackrel{\rm def}{=}}
\newcommand{\non}{\nonumber}
\newcommand{\bquo}{\begin{quote}}
\newcommand{\enqu}{\end{quote}}
%%%%%%%%%%%%%%%%
\renewcommand{\(}{\begin{equation}}
\renewcommand{\)}{\end{equation}}
%%%%%%%%%%%%%%%%%%%%%%%%%%%%%%%%%% definitions
\def \eqn#1#2{\begin{equation}#2\label{#1}\end{equation}}
\def\IZ{{\mathbb Z}}
\def\IR{{\mathbb R}}
\def\IC{{\mathbb C}}
\def\IQ{{\mathbb Q}}
\def\de{\partial}
\def\Tr{ \hbox{\rm Tr}}
\def\H{ \hbox{\rm H}}
\def\HE{ \hbox{$\rm H^{even}$}}
\def\HO{ \hbox{$\rm H^{odd}$}}
\def\K{ \hbox{\rm K}}
\def\Im{ \hbox{\rm Im}}
\def\Ker{ \hbox{\rm Ker}}
\def\const{\hbox {\rm const.}}
\def\o{\over}
\def\im{\hbox{\rm Im}}
\def\re{\hbox{\rm Re}}
\def\bra{\langle}\def\ket{\rangle}
\def\Arg{\hbox {\rm Arg}}
\def\Re{\hbox {\rm Re}}
\def\Im{\hbox {\rm Im}}
\def\exo{\hbox {\rm exp}}
\def\diag{\hbox{\rm diag}}
\def\longvert{{\rule[-2mm]{0.1mm}{7mm}}\,}
\def\a{\alpha}
\def\dag{{}^{\dagger}}
\def\tq{{\widetilde q}}
\def\p{{}^{\prime}}
\def\W{W}
\def\N{{\cal N}}
\def\hsp{,\hspace{.7cm}}

\def\br{\nonumber\\}
\def\IZ{{\mathbb Z}}
\def\IR{{\mathbb R}}
\def\IC{{\mathbb C}}
\def\IQ{{\mathbb Q}}
\def\IP{{\mathbb P}}
\def \eqn#1#2{\begin{equation}#2\label{#1}\end{equation}}

\newcommand{\C}{\ensuremath{\mathbb C}}
\newcommand{\Z}{\ensuremath{\mathbb Z}}
\newcommand{\R}{\ensuremath{\mathbb R}}
\newcommand{\rp}{\ensuremath{\mathbb {RP}}}
\newcommand{\cp}{\ensuremath{\mathbb {CP}}}
\newcommand{\vac}{\ensuremath{|0\rangle}}
\newcommand{\vact}{\ensuremath{|00\rangle}                    }
\newcommand{\oc}{\ensuremath{\overline{c}}}
\begin{titlepage}
\begin{flushright}
ULB-TH/09-06\\
%hep-th/yymmnnn\\
\end{flushright}
\bigskip
\def\thefootnote{\fnsymbol{footnote}}

\begin{center}
{\Large
{\bf
M2-brane Flows and the Chern-Simons Level\\
\vspace{0.1in}
%and Blown up Six Cycles \\
%\vspace{0.1in}
%M2-brane RG Flows
}
}
\end{center}

\bigskip
\begin{center}
%{\large
{Chethan
KRISHNAN$^1$\footnote{\texttt{Chethan.Krishnan@ulb.ac.be}}, Carlo MACCAFERRI$^1$\footnote{\texttt{cmaccafe@ulb.ac.be}} and Harvendra SINGH$^2$\footnote{\texttt{h.singh@saha.ac.in}}}
%}
\\
\end{center}

\begin{center}
\vspace{1em}
{$^1$\em  { International Solvay Institutes,\\
Physique Th\'eorique et Math\'ematique,\\
ULB C.P. 231, Universit\'e Libre
de Bruxelles, \\ B-1050, Bruxelles, Belgium\\}}
\end{center}

%\begin{center}
%{\large Harvendra SINGH$^2$\footnote{\texttt{h.singh@saha.ac.in}}} \\
%\end{center}
\begin{center}
{$^2$\em  { Theory group, Saha Institute of Nuclear Physics,\\
1/AF, Bidhannagar, Kolkata 700064, India\\}}

\renewcommand{\thefootnote}{\arabic{footnote}}

\end{center}

\noindent
\begin{center} {\bf Abstract} \end{center}
The Chern-Simons level $k$ of ABJM gauge theory captures the orbifolding in the dual geometry. This suggests that if we move the membranes away from the tip of the orbifold to a smooth point, it should trigger an RG flow that changes the level to $k=1$ in the IR. We construct an explicit supergravity solution that is dual to this shift from generic $k$ to $k=1$. In the gauge theory side, we present arguments for why this shift is plausible at the end of the RG flow. We also consider a resolution of the orbifold for the case $k=4$ (where explicit metrics can be found), and construct the smooth supergravity solution that interpolates between $AdS_4 \times S^7/\IZ_4$ and $AdS_4 \times S^7$, corresponding to localized branes on the blown up six cycle. In the gauge theory, we make some comments about the dimension four operator dual to the resolution as well as the associated RG flow. %In both the resolved and unresolved cases, we discuss aspects of the RG flows on the two sides of the duality.

%We construct a supergravity solution that interpolates between $AdS_4\times S^7/\IZ_4$ (in the UV) to $AdS_4\times S^7$ (in the IR) as we approach a stack of M2-branes localized at a point on the resolution of the $\IC^4/\IZ_4$ orbifold. This corresponds, in the dual three dimensional gauge theory, to giving a VEV which triggers an RG flow that effectively reduces the Chern-Simons level of the ABJM gauge theory to $k=1$.

\begin{center}
\end{center}

%\begin{center}
\vspace{1.6 cm}
\vfill

\end{titlepage}
\hfill{}
\bigskip

\tableofcontents

\setcounter{footnote}{0}
\section{Introduction}

\noindent
The near horizon limit of the membrane solution in 11-D supergravity gives rise to a background which is asymptotically $AdS_4\times S^7$ (see e.g., \cite{Stelle} for the solitonic p-brane solutions in various supergravity theories). By gauge-gravity duality, we expect that this gravitational background also has a dual description in terms of the worldvolume gauge theory on M2-branes. One of the interesting developments in the past year has been the explicit construction of these gauge theories. Generalizing the pioneering work of Bagger, Lambert \cite{BL} and Gustavsson \cite{G}, superconformal gauge theories living on $N$ membranes probing the $\IC^4/\IZ_k$ orbifold singularity have been identified (``ABJM theory") \cite{ABJM}\footnote{The flat space limit corresponds to the special case $k=1$.}. This construction has also been further generalized to include some other classes of toric singularities in, e.g., \cite{Amihay}.

One interesting feature of ABJM theory is that they are Chern-Simons gauge theories coupled to matter, and they are characterized by the level $k$ which defines the orbifold on the gravity side. The Chern-Simons level is effectively the coupling of the theory, implying that the flat space limit corresponds to a strongly coupled gauge theory. This is unlike the familiar case of $AdS_5/CFT_4$, where the gauge-coupling is a free parameter in flat space.

If we move the branes away from the orbifold, the membranes are locally in flat space. In the near horizon limit, this will translate to the emergence of an $AdS_4\times S^7$ throat. The gauge theory on the worldvolume of membranes in flat space is ABJM at level $k=1$. The interpolation on the gravity side (Note that the solution is still singular in this case because the geometry is singular.) between $AdS_4 \times S^7/\IZ_k$ and $AdS_4 \times S^7$ should correspond to giving an appropriate vev that triggers an RG flow in the gauge theory picture. This RG flow should lead us from generic $k$ to $k=1$ ABJM. In practice, giving a VEV changes ABJM theory to ${\cal N}=8$ SYM theory which is not conformal, but the expectation is that it runs in the IR to the correct ABJM theory. We present some insights into how this happens. In particular, we construct the explicit supergravity solution that exhibits this shift.

Another related problem we consider is the resolution of the orbifold. We can construct explicit metrics on the resolution when $k=4$. In this case, the resolution is a six-cycle. We consider stacks of branes on the resolution, which looks far away in the UV like $AdS_4 \times S^7/\IZ_4$ and construct a smooth interpolation between that and the usual $AdS_4 \times S^7$. The solution is fully non-singular. It is expressed in terms of angular harmonics on $\IC\IP^3$, and the sum over the angular harmonics reproduces the AdS throat close to the stack.

\begin{figure}
\begin{center}
\includegraphics[%width=0.9\textwidth,
height=0.35\textheight
]{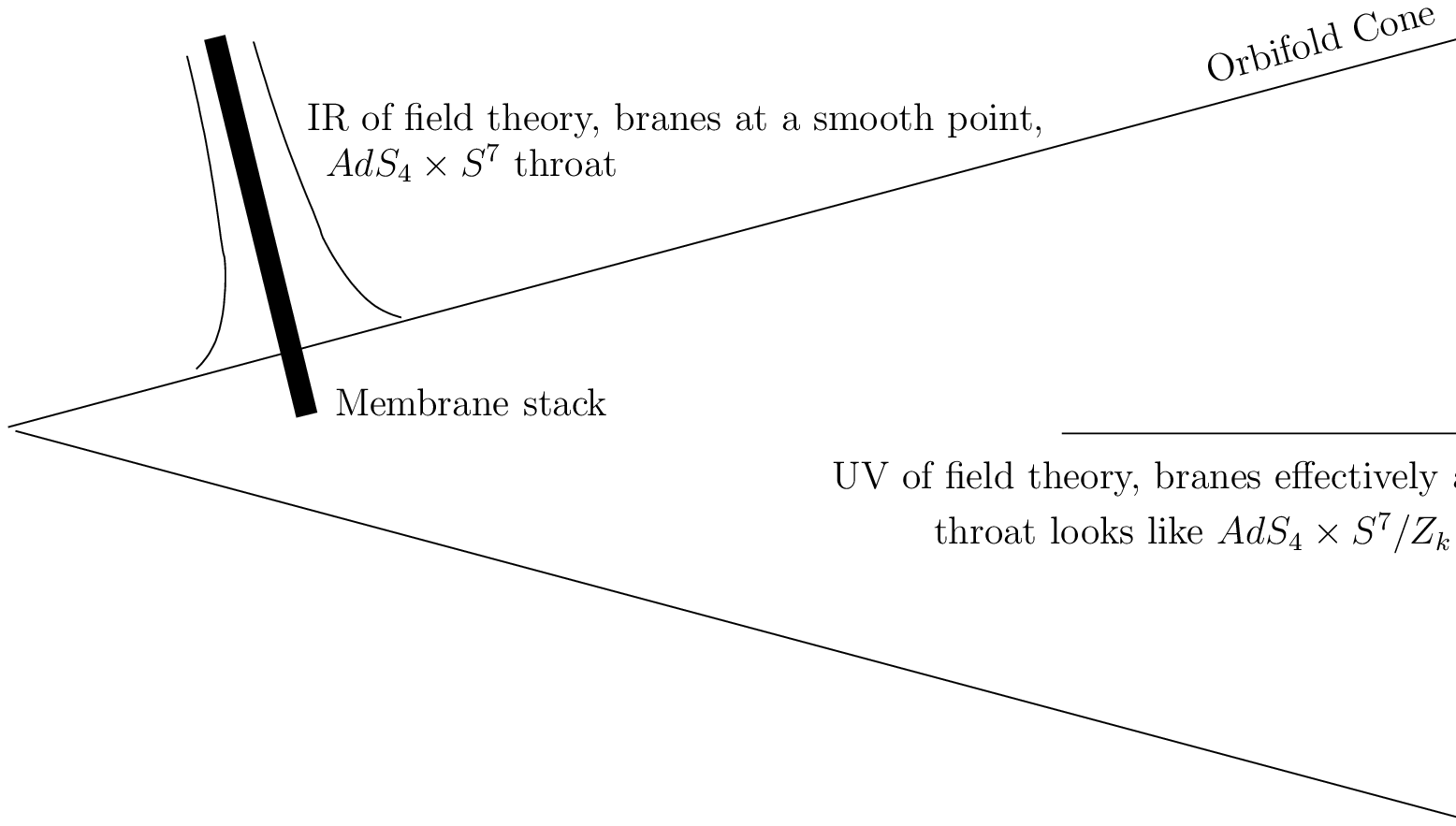}%\input{pen.pstex_t}120 390 640 800
\caption{Schematic holographic picture of the RG Flow between two ABJM theories with generic $k$ (UV), and $k=1$ (IR). This is the unbackreacted geometry, even though we have indicated the location of the emergent $AdS$ throat.}
\label{toric}
\end{center}
\end{figure}

The basic message of the paper is that there exist quantum field theories that flow from a UV fixed point which is a Chern-Simons-matter theory at level $k$ to an IR fixed point which is a Chern-Simons-matter theory at level $k=1$. This is a general consequence of the ABJM construction. Our supergravity solutions, as well as the Abelian RG flow and the moduli space arguments on the gauge theory side, are presented as evidence for this. It should be emphasized that the level shift that we consider is very different from the familiar perturbative one-loop shift in the level that arises in Chern-Simons theories. Our level shift arises at the end of an RG flow in the IR and is not visible by doing perturbation theory around the UV fixed point. We present a schematic diagram of the ``holographic" version of the RG flow in Figure 1. We show only the unresolved case, the ``tip" will of course be smooth for the resolution.

%there exists RG flows that relate various Chern-Simons-matter theories with distinct levels.

In the next section, after reviewing ABJM theory, we consider supergravity solutions with brane sources on the orbifold and their holographic RG flows. In section 3, we discuss the resolution of the orbifold and branes on them, including the emergence of the AdS throat near the stack. Section 4 considers the gauge theory aspects. Finally we conclude with some comments.

Interpretational aspects of Bagger-Lambert theory were clarified in \cite{Tong, int}, the 3-algebra structures underlying BLG and ABJM were discussed in \cite{Cherkis}, various formulations of membrane theories was considered in \cite{superspace}, integrability in $AdS_4/CFT_3$ were discussed in \cite{integrability}\footnote{It has been argued in \cite{Dmitri} that the the supersymmetric sigma model for the string in $AdS_4 \times \IC \IP^3$ is most appropriately thought of as arising from dimensional reduction. In this framework, the theory that results is a ``twisted" supercoset, and the classical integrability is not clear.}, \cite{Krishnan} and general classes of three-dimensional CFTs and their gravitational duals were constructed in \cite{cft3, ABJ}. An approach to membranes using rank two tensor fields (without 3-algebras) was proposed in \cite{HS}. While this work was being completed, \cite{Warner} appeared, which also considers M2-brane flows, but in a different context. Chern-Simons level shifts in the context of holographic condensed matter systems have been investigated in \cite{Fujita}.

\section{The Orbifold}

In this section we present some of the relevant details of gauge-gravity duality on $\IR^{2,1} \times \IC^4/\IZ_k$. %, with emphasis on the special case we are interested in, namely $k=4$.
We start by reviewing some relevant aspects of ABJM gauge theory.

\subsection{ABJM Gauge Theory}

ABJM theory is a $2+1$ dimensional ${\cal N}=6$ superconformal model with $U(N) \times U(N)$ gauge-fields $(A, \hat A)$, whose action takes the form of a $(k,-k)$-level Chern-Simons theory. The gauge-fields are coupled to two sets of two bifundamental scalars $(Z^A, W_A)$, $A= 1,2$. Here $Z^A$ transform in the $({\bf N, \bar N})$ and $W_A$ in the $({\bf \bar N, N})$ representation\footnote{Note that the indices on the fields are placed so that the complex conjugate of $W$ transforms like $Z$.}. The theory also contains superpartners of these fields, some of which are auxiliary fields that can be integrated out. The final form of the action in component fields is presented in \cite{Benna}, we will just right down the kinetic terms for the bosons here:
\bea
S=\int d^3 x\Big[ \frac{k}{4\pi}\epsilon^{\mu\nu\lambda}\Big({\rm Tr}\big(A_\mu\partial_\nu A_\lambda+\frac{2i}{3}A_\mu A_\nu A_\lambda\big)- {\rm Tr}\big(\hat A_\mu\partial_\nu \hat A_\lambda+\frac{2i}{3}\hat A_\mu \hat A_\nu \hat A_\lambda\big)\Big)\nonumber\\ -{\rm Tr}(D_\mu Z)^\dagger D^\mu Z- {\rm Tr}(D_\mu W)^\dagger D^\mu W + {\rm \ fermionic \ and \ potential \ terms\Big].}
\eea
The traces are in the appropriate representations of the relevant groups and the covariant derivatives are defined by
\bea
D_\mu Z &=& \partial_\mu Z-i Z \hat A_\mu+iA_\mu Z, \\
D_\mu W &=& \partial_\mu W+i \hat A_\mu W-i W A_\mu.
\eea
On general grounds \cite{Deser}, one expects that the Chern-Simons level is quantized to be integer valued when the gauge group rank is $\ge 2$. The theory is weakly coupled when $k$ is large, and the 't Hooft coupling takes the form $N/k$.

One of the general expectations of the gauge-string correspondence is that the moduli-space of vacua of the gauge theory is dual to the geometry on which the branes move. Indeed, one piece of evidence that ABJM theory is indeed the correct theory for $N$ M2-branes probing a $\IC^4/\IZ_k$ singularity, is that the moduli space is precisely $(\IC^4/\IZ_k)^N/S_N$. We will use this fact when we discuss RG flows in section 4, for the moment, we merely mention that this connection is made through the identification $C^I=\{Z^A, W^{B\dagger} \}$, see footnote 3. With $I$ running from 1 through 4, it can be shown that $C^I$ correspond to the complex coordinates $\IC^4$, up to a discrete $\IZ_k$ identification. This notation manifests the $SU(4)_R$ invariance of the theory.

\subsection{Membrane Supergravity}

The space $\IC^4/\IZ_k$ is defined as the four-dimensional complex space $\IC^4$ after the identification
\eqn{orbifold}
{ ( w_1, w_2, w_3, w_4 ) \sim ( e^{2\pi i/k} w_1,e^{2\pi i/k} w_2, e^{2\pi i/k} w_3 ,e^{2\pi i/k} w_4 ).}
These $w_i$'s are related to the gauge theory variables as $C^I \sim  \{w_a, \bar w_b\}$, i.e., the natural complex structure of ABJM is defined in terms of $\{ w_1, w_2, \bar w_3, \bar w_4\} $, which preserves ${\cal N}=6$  supersymmetry for orbifolding by any $k$. For us, the complex structure will often merely be a crutch for writing explicit metrics. Note that the complex structure we are defining is also preserved under the orbifolding for $k=1, 2$ and $4$. This fact is of not particular importance in the unresolved case, but when we construct smooth solutions on the resolution of $\IC^4/\IZ_4$ later, this will make sure that the solution has some supersymmetry and is stable.

The flat metric on $\IC^4$ descends to the cone metric on the orbifolded space, with the origin as the orbifold singularity:
\eqn{orb}{ds_8^2=dr^2+r^2d\Omega_{S^7/\IZ_k}^2}
We are interested in M2-branes in the background $\IR^{2,1} \times \IC^4/\IZ_k$. A stack of M2-branes in a background acts as a source for the 11 dimensional supergravity equations of motion. A standard ansatz for solving these equations of motion can be found for example, in \cite{Stelle}. The non-vanishing fields take the form:
\begin{eqnarray}\label{m2ansatz}
ds^2&=&H^{-2/3}(y)\eta_{\mu\nu}dx^{\mu}dx^{\nu}+H^{1/3}(y)ds_8^2,
\hspace{0.3in} \\
F_4&=&dH^{-1}\wedge dx^0 \wedge dx^1 \wedge dx^2,
\end{eqnarray}
The $ds_8^2$ piece in the
metric denotes the dimensions transverse to
the M2-branes, and is given in our case by (\ref{orb}). The worldvolume
Minkowskian metric of the M2-branes is (2+1)-dimensional, and the entire solution is
captured by a single function, $H(y)$, where $y$ denotes the
coordinates on the transverse space. This function is called the warp
factor and it fully describes the solution. With this ans\"atz, the supergravity EOMs (with source terms
for the branes) reduce to just one equation, the Green's equation on the transverse space ($y_0$ denotes the
location of the stack):
\eqn{gr}{\Box_y H(y,y_0)=-\frac{C}{\sqrt{g_8}}\delta^8(y-y_0), \ \ {\rm with} \ \
C=2\kappa_{11}^2T_2N.}
where we denote the determinant of the 8-metric by $g_8$. The
strength of the source is captured by $C=2
\kappa_{11}^2T_2N$ where $T_2$ is the brane tension
and $\kappa_{11}$ is the 11-D Newton's constant. %{\bf (Maybe we should relate these to the definition of $l_p$ in Harvendra's previous paper.)}.

For the case of the unresolved $\IC^4/\IZ_k$, when we place the stack at
the orbifold singularity at the apex of the cone, this equation can be immediately solved because
the warp factor depends only on the radial coordinate. Green's
equation takes the form
\eqn{Gr1}{\frac{1}{r^7}\frac{\partial}{\partial
r}\Big(r^7\frac{\partial}{\partial r} H\Big)= -\frac{3 k C}{\pi^4
r^7}\delta(r).}
The normalization of the delta function accounts for the fact that we are
ignoring the angular dependence. It arises from an integral over the
angles. This is analogous to the fact that in 3-dimensions, if we
are looking at sources at the origin ($r_0=0$), then we can replace
$\frac{1}{r^2\sin
\theta}\delta(r-r_0)\delta(\theta-\theta_0)\delta(\phi-\phi_0)
\rightarrow \frac{1}{4\pi r^2} \delta(r)$
when dealing with test-functions that are sufficiently well-behaved at
the origin. The $4 \pi$
here arises through an angle integral as well: $\int_0^{\pi} \sin \theta
d\theta\int_0^{2\pi}
d\phi = 4\pi$.

Away from
the origin the equation is easily integrated, and integrating over the
delta function fixes the constant of integration:
\eqn{warp1}{H(r)=\frac{L^6}{r^6} \ \ {\rm where} \  \ L^6=\frac{k C}{2 \pi^4}.}
Notice that in integrating the Poisson equation, we assume that the Green's function falls off to zero at infinity. This is tantamount to assuming that we are in the near horizon region. If we allow a non-zero constant at infinity in $H(r)$, we will have membranes in an asymptotically flat space.
With the $H(r)$ obtained above, if we do the substitution $z=L^3/2r^2$, we end up with
\eqn{S5z3}{ds^2=
\frac{L^2}{4z^2}(dz^2+\eta_{\mu\nu}dx^{\mu}dx^{\nu})+L^2d\Omega_{S^7/\IZ_k}
=\frac{L^2}{4}ds^2_{AdS_4}+L^2d\Omega_{S^7/\IZ_k}}
which is nothing but $AdS^4\times S^7/\IZ_k$ with appropriate radii.

Another thing that we could do is to put the stack away from the tip, in
which case we expect that far away, the solution should still look like
the one we found above. But close to the stack, now we should see the
emergence of the AdS throat because the stack is now at a smooth point.
The full solution will have a singularity at $r=0$. We check these expectations in the next subsection. Moving the stack away from the origin is equivalent to turning on a VEV in the gauge theory. This triggers an RG flow and the near-horizon limit will correspond to the IR fixed point of this flow. Aspects of the gauge theory side will be discussed in section 4.

\subsection{Membranes Away From the Orbifold}

Here we compute the Green's function for a D-brane stack on
the unresolved orbifold, but away from the tip. We will call the radial coordinate $\rho$ instead of $r$. This is for ease of comparison with the resolved case we will consider later on. The resolution parameter $a=0$ in the present case. The stack of branes will be placed at $\rho_0 \ne 0$, away from the tip.
%($\rho\equiv r$ and $a=0$ since the space is unresolved, but $\rho_0 \ne 0$, away from the tip.).
Far away
from the stack, we expect
to reproduce the behavior we calculated in section 2.2, but we also
expect to see the AdS throat in the near-horizon region because the stack
is no longer at a singular point. The fact that the space is unresolved
will be reflected in the fact that the solution is still singular.

To simplify the computations, without loss of generality we will look at the case where the stack is
at the point $\rho=\rho_0, \ \xi=0, \beta=0$. The location
$\xi=0$ kills the dependence of the Green function on the other angles
of the $\IC\IP^3$ \footnote{See section 3.2 for a more complete discussion. We are using $S^7$ as a circle fibration over $\IC\IP^3$. The $\IC\IP^3$ metric is presented in an appendix.}.
We want to retain the dependence on $\beta$ because otherwise our solution will correspond to branes smeared over the fibered circle. So we will look for solutions of the form
$H(\rho,\xi, \beta)$.
%\footnote{It turns out that the more general case of $\psi$-dependent warp factor can also be solved exactly in terms of Jacobi polynomials (this time, with more general quantum numbers than what we saw in Section 4), but we will not present the solution here.}
The form of the
Laplacian permits such a choice. The equation to be solved takes the form
\begin{eqnarray}
\Box_\rho H+\frac{1}{\rho^2}\Big(\Box_{\xi}H+\frac{16}{\cos^2 \xi}\partial_\beta^2 H\Big)=-\frac{ C}{\pi^3\rho^7\sin^5
\xi \cos \xi}\delta(\rho-\rho_0)\delta(\xi)\delta(\beta),\\
{\rm where}\ \  \Box_\rho=\frac{1}{\rho^7}\frac{\partial}{\partial
\rho}\Big(\rho^7\frac{\partial}{\partial \rho}\Big),\hspace{1.5in}
%\ \ {\rm and} \ \
%\Box_{\sigma,\psi}= \Box_\sigma+\frac{9}{\cos^2 \sigma}\partial_\psi^2, \
\ \ \nonumber
\end{eqnarray}
with $\Box_\xi$ as defined by (\ref{ang}). We first solve the azimuthal (fiber) part
\bea
\partial_\beta^2 \chi_m (\beta)=-m^2 \chi_m(\beta) \ {\rm  to \ find}\ \chi_m(\beta) =\sqrt{\frac{k}{2\pi}} e^{i m \beta}.
\eea
Note that $m$ runs over multiples of $k$ so that the function is periodic. The normalization is also affected by the quotienting of the fiber. Plugging this into the rest of the angular part, we find
\begin{eqnarray}
&\Big(\Box_\xi+2l (2l+6)-\frac{16 m^2}{\cos^2 \xi} %-\frac{36 m^2}{\cos^2 \sigma}
\Big)\Sigma_l(\xi)=0,&
%&\big(\partial_\psi^2+4m^2)\Psi(\psi)=0&
\end{eqnarray}
The eigenvalue $l$ has been defined for future convenience (the solution is regular everywhere in its range only if $l$ is integral). The solutions of the above equations are in terms of Jacobi polynomials, see appendix.
The full normalized angular solutions $Y_{l,m}\equiv \Sigma_l(\xi) \times \chi_m(\beta)$ are
\eqn{ylm}{Y_{l,m}(\xi,\beta)=%\sqrt{\frac{k}{2\pi}}
\sqrt{\frac{2(2l+3)(l+2m+2)(l+2m+1)}{(l-2m+2)(l-2m+1)} }%(l+3m+1)}{2\pi(l-3m+1)}}
%\frac{\Gamma(l+3m)}{\Gamma(6m)\Gamma(l-3m)}
%\cos^{6m}(\sigma)
\cos^{4m} \xi \ P_{l-2m}^{(4m,2)}(-\cos 2\xi)\sqrt{\frac{k}{2\pi}}e^{i m \beta}%\exp(2im \psi)
.}
The fact that $l$ is integral, $l\ge 0$, and that $m$ is integral with $|2m| \le l$ can be obtained from the appropriate periodicity in $\beta$ and the regularity of the harmonic functions at the poles. This is exactly analogous to the familiar case of harmonics on a 2-sphere.
%For the moment, we will not dwell on the range of $m$ etc., because these will not be important to us. %The only eigenfunction whose explicit value we will need is the $l=0, m=0$ case.
%The $P_n^{(l,m)}(x)$ are again Jacobi Polynomials. Among other things, we
%have used orthonormality of Jacobi Polynomials
%\cite{Handbook} to fix normalizations.
%Also, $l, m$ get constrained to be integral in the process. Notice that
%the angular wave functions reduce to the ones obtained in (\ref{angsol}),
%including the normalization, when we set $m=0$ (and integrate over $\psi$
%to get rid of a factor of $2\pi$ to compensate for the
%difference in delta-function normalizations in the two cases.).	

Now we turn to the radial part of the equation. It looks like
\eqn{apprad}{\frac{\partial^2 H_l}{\partial
\rho^2}+\frac{7}{\rho}\frac{\partial H_l}{\partial \rho}-\frac{2l
(2l+6)}{\rho^2} H_l=-\frac{ C}{\pi^3\rho^7}\delta(\rho-\rho_0),}
whose solution, after matching the function and its derivative through
the delta function is
\begin{eqnarray}
H_l(\rho,\rho_0)=
\left\{ \begin{array}{ll}
  \displaystyle{\frac{ C}{2 \pi^3(2l+3)} \rho_0^{-(2l+6)}\rho^{2l} }&
\ \ \rho
\leq
\rho_0, \\
 \\
  \displaystyle{\frac{ C}{2 \pi^3(2l+3)} \rho^{-(2l+6)}\rho_0^{2l} }
& \ \ \rho
\geq
\rho_0.\\
\end{array} \right.
\end{eqnarray}

The full solution looks like
\eqn{fin}{H(\rho,\rho_0;\xi,0;\beta,0)=\sum_{l,m}
H_l(\rho,\rho_0)Y_{l,m}^*(0,0)Y_{l,m}(\xi,\beta).}
Restricting to $l=0$ (which forces $m=0$) and using properties of Jacobi polynomials gives the dependence far away from the source
\eqn{faraway}{H(\rho) \rightarrow \frac{ C}{6 \pi^3 \rho^6}\times
6 \times \frac{k}{2\pi}=\frac{k C}{2 \pi^4\rho^6}.}
This reproduces the result obtained in eqn. (\ref{warp1}),
where the stack was
assumed to be at the tip.

\subsection{Emergence of the AdS throat}

We will be rather quick in this subsection because essentially the same ideas, in a somewhat more complicated form, will re-appear in the resolved case: we present more detail there. The reader might wish to come back to this section after reading section 3.3.

To see the emergence of the $AdS$ throat near the stack, we first expand the radial equation of motion (\ref{apprad}) near the source: $\rho=\rho_0+\tilde\rho$, and retain only first order pieces. The homogeneous part is
\bea
\frac{\partial^2 H_l}{\partial
\tilde\rho^2}+\frac{7}{\rho_0}\frac{\partial H_l}{\partial \tilde\rho}-\frac{2l
(2l+6)}{\rho_0^2} H_l=0.
\eea
This is easily solved, and the solution that falls off far away is
\bea
H_l=A_l \exp\Big\{-\frac{\Big(7+\sqrt{49+16l(l+3)}\Big)\tilde \rho}{2\rho_0} \Big\}
\eea
The normalization $A_l$ must be fixed by integrating across the source according to
\bea
-\frac{C}{\pi^3}=\left(\rho^7\frac{dH_l}{d\rho}\right)\Big|_{\tilde\rho=0}.
\eea
We will only need the dependence on $l$ (for large $l$) and $\tilde \rho$, so we will write the final result as
\bea
H_l \sim \frac{\exp(-\alpha l \tilde \rho)}{l} %e^{-\alpha l r}
\eea
for some positive $\alpha$ that does not contain $l$ or $\tilde \rho$.

Our aim is to show that the warp factor reduces to the hexic $AdS$ form close to the stack. To demonstrate this, it is easiest to approach the stack along $\beta=0, \xi=0$, in which case (\ref{fin}) reduces to
\bea
H &\sim& \sum_{l,m} (2l+3) (l+2m+2)(l+2m+1)(l-2m+2)(l-2m+1) H_l \nonumber\\
&\sim & \sum_l l^6 H_l \sim \sum_l l^5\exp(-\alpha l \tilde \rho),
\eea
where we have used the Jacobi polynomial identity $P_{l-2m}^{(4m,2)}(-1)=(-1)^l(l-2m+2)(l-2m+1)/2$ and also the fact that $|2m|\le l$. The final expression can be thought of as a regulated sum and we can write it as an integral (see the discussion around (\ref{www2}) for an explanation):
\bea
H \sim \int_0^{1/\tilde \rho} l^5 \exp(-\alpha l\tilde \rho) dl \sim \frac{1}{\tilde \rho^6}\int_0^1 x^5 \exp(-\alpha x) dx \sim \frac{{\rm const.}}{\tilde \rho^6},
\eea
which is the hexic fall-off necessary for the emergence of the $AdS_4 \times S^7$ throat (cf. \ref{warp1}, \ref{S5z3}). It is necessary to note here also that around the smooth point $\rho_0$, along the line of approach to the stack that we are considering, the metric (\ref{orb}) can be written in flat form with a radial coordinate $\tilde \rho$. The easiest way to notice this is to note that away from the orbifold singularity, the metric is flat to begin with.

%close to the stack is analogous to the resolved case, so we will discuss it in section 3.

\section{M2-branes on the Resolution}

The supergravity solution that we constructed in the previous section  is singular, even when the stack is moved away from the orbifold, because the space-time is not smooth. It would be interesting to construct a solution on a fully smooth geometry and that is what we set out to do in this section. Later, we also make some comments about the dual gauge theory interpretation of this resolution, and the RG flows on them.

Before we start, we emphasize that the ``resolution" that we are after is merely a way to have a smooth deformation of the cone where we have a way to construct an explicit solution. In particular, the use of the complex structure is that in the next subsection we will be able to use its existence to derive a form for the metric on the resolution. It does not have a simple map to the gauge theory moduli space, unlike in $AdS_5/CFT_4$. For $k=4$ the supersymmetry is preserved, ensuring that the solution is stable \footnote{See also the discussion at the beginning of Section 2.2. We thank S. Kuperstein for discussions on this.}. 

\subsection{The Resolved Geometry}

We start with some general comments about $\IC^n/\IZ_n$ orbifolds and their resolutions. The symmetries of the $\IC^n/\IZ_n$ orbifold are sufficiently restrictive that demanding that the resolved metric respect these symmetries (along with the fact that it is Ricci flat and K\"ahler), completely fixes it. When $n=2$, this gives rise to the familiar Eguchi-Hanson ALE space, and when $n>2$ this gives us a simple way to construct gravitational instantons in higher dimensions.

The $\IZ_n$ orbifold action is a discrete subgroup of the $SU(n)$ isometry which rotates the various $w$'s (see (\ref{orbifold}) for the $n=4$ case). We can take the K\"ahler form
$K(r)$ on this space to depend only on $r^2=\sum_i^n|w_i|^2$.
Since the space is Calabi-Yau, among other things, it is both
Ricci-flat and K\"ahler. So the metric can be written as $g_{i\bar
j}=\partial_i \partial_{\bar j} K$,
and then the Ricci-flatness condition turns out to be a
differential equation for $K(r)$:
\begin{eqnarray}
{\rm det}(\partial_i\partial_{\bar j}K)={\rm const.}
\end{eqnarray}
From the explicit form of the matrix, it can be seen that this reduces (after absorbing the irrelevant constant by re-scaling $K(r)$) to
\eqn{K}{(K')^{(n-1)}(r^2K''+K')=1.}
The primes here are with respect to $r^2$. It will prove convenient to
introduce a new function ${\cal F}$ defined by
\eqn{F}{{\cal F}\equiv r^2 K',}
in terms of which the differential equation above has the simple solution
\eqn{F1}{{\cal F}=(r^{2n}+a^{2n})^{1/n},}
with $a^{2n}$ an integration constant which reflects the resolution of the
space. By tuning $a$ to zero, we can recover the unresolved space. It is
possible to integrate ${\cal F}$ once again to express
$K(r)$ in terms of hypergeometric functions. We present it here for completeness:
\bea
\label{kahler}
K(r)=r^2 \ {}_2F_1\Big(-\frac{1}{n},-\frac{1}{n};1-\frac{1}{n};-\frac{a^{2n}}{r^{2n}}\Big)
\eea

So far, everything we said is valid for any $n$. Our real interest is in the special case when the transverse space is 8-dimensional and corresponds to membrane theories, so now we specialize to the case of $n=4$. This $\IC^4/\IZ_4$ orbifold will be a central object in this section.

With the K\"ahler potential at hand, now we can define some convenient angular variables to write down an explicit form of the metric. We define eight real coordinates through\footnote{Some hints for the choice of this parametrization can be found from the Fubini-Study metric on $\IC\IP^3$. See appendix A.}
\bea\label{compcord}
w_1&=&r\ \sin \xi\ \sin \alpha \ \sin {\theta \over 2}\
e^{i{(\psi-\varphi)\over 2}}\ e^{i{\beta\over 4}}\ e^{i{\chi\over 2}} \\
w_2&=&r\ \sin \xi\ \sin \alpha \ \cos {\theta \over 2}\
e^{i{(\psi+\varphi)\over 2}}\ e^{i{\beta\over 4}}\ e^{i{\chi\over 2}} \\
w_3&=&r\ \sin \xi\ \cos \alpha \ e^{i{\beta\over 4}}\
e^{i{\chi\over 2}} \\
w_4&=& r\ \cos \xi\ e^{i{\beta\over 4}}\label{compcord4}
\eea
Using this definition of $w_i$, we can calculate the metric on the resolution directly as $ds^2=g_{i\bar j}dw^i
d{\bar w}^{\bar j}$, with $g_{i\bar j}=\partial_i \partial_{\bar j} K$.
The result, once the dust settles, is
\begin{eqnarray}
ds^2={\cal F}'dr^2+\frac{{\cal
F}'r^2}{16}(d\beta-A)^2 +{\cal F} \ ds^2_{\IC\IP^3},
\end{eqnarray}
where the $\IC\IP^3$ metric is as defined in the appendix, and $A$ is as given by (\ref{k=4}), with $k=4$.
 Now we can define a new radial coordinate through $\rho^2 \equiv {\cal F}$, and we reach the simple and useful form:
\bea \label{metricr}
ds^2={d\rho^2 \over \Big(1- {a^8 \over \rho^8}\Big)}+{\rho^2 \over 16}\Big(1-{a^8 \over \rho^8}\Big)(d\beta -A)^2+ \rho^2 ds^2_{\IC\IP^3}.
\eea
Notice that this structure is an immediate generalization of Eguchi-Hanson, thought of as a resolution of $\IC^2/\IZ_2$ with the singularity replaced by a two-cycle $\IC\IP^1$. Here the resolution is a six-cycle. A purely algebraic proof of this fact is presented in Appendix D. The $\beta$ corresponds to the $U(1)$ fibration over $\IC\IP^3$.

%{\bf If we do explicit angles, then we should spell out the $A$ and the CP3 metric using those angles.}

\subsection{Poisson Equation on the Resolution}

Our aim is to construct the supergravity solution generated by a stack of M2-branes on this resolved space.

We will first consider the case where we put the stack of M2-branes smeared over the resolution of the orbifold, so that the we can make the simplifying assumption that the warp factor is only a function of the radial coordinate. This was done in \cite{harvendra}. In the case of the resolved conifold, an analogous computation was done originally in \cite{tseytlin}. The (homogeneous part of the) equation to be solved in our case can in fact be written in the form\footnote{See for instance equation (\ref{sm5}). The smeared Laplacian is just the radial part of the full Laplacian.}
\bea
\frac{1}{\rho^7}\partial_\rho\Big(\rho^7\big(1-\frac{a^8}{\rho^8}\big)\partial_\rho H\Big)=0.
\eea
Since this is effectively a first order equation, it can be solved by direct integration. The delta-function can be used to determine the overall constant. More directly, we can also fix it by comparing with (\ref{warp1}) as $\frac{\rho}{a}\rightarrow \infty$. The end result is
\bea\label{smeared}
H^{{\rm smeared}}=-\frac{3C}{\pi^4a^6}\Big[ {1 \over 2}\log\Big(\frac{\rho^2-a^2}{\rho^2+a^2}\Big)+{\pi \over 2}-\tan^{-1}\Big(\frac{\rho^2}{a^2}\Big)\Big]
\eea

The full nonsingular M2-brane solutions on the resolved $\IC^4/\IZ_4$
can be constructed by allowing the brane-stack to be localized at
particular
angular location on
the ``blown-up" $\IC\IP_3$ instead of homogeneously distributing them over the
$\IC\IP_3$. This is an approach adopted by Klebanov-Murugan for
obtaining regular D3-brane solutions over the resolution of the conifold singularity
\cite{KM}, where the resolution was a two-cycle. The same method was also  adopted in \cite{KK} to write down regular
D3-brane solutions over the resolved $\IC^3/\IZ_3$ orbifold geometry. Related work can be found in \cite{KK2,Hin}.

We will put the stack at $\rho=a$, where the fibration has shrunk to zero size. This means that our warp factor will no longer depend on $\beta$. Also (without loss of generality) we will put the M2-brane sources at $\xi=0$, where the rest of the cycles of $\IC\IP^3$ collapse to zero size (see the $\IC\IP^3$ metric presented in the appendix.). This means that $H$ does not depend on the rest of the angles of $\IC\IP^3$ as well. So we can look for a warp factor in the form $H(\rho,\xi)$.

%{\bf If we can write down an explicit angular parametrization, then we can make some more precise comments. But everything we want can be determined one way or another, so I haven't tried to do it.}

We make a radial-spherical ansatz of the following type for the Green function:
\eqn{Hgeneric}{H(\rho,\rho_0=a,\xi,\xi_0=0)=\sum_l
H_l(\rho,\rho_0=a)Y_l^*(\xi_0=0)Y_{l}(\xi).}
What makes this possible is the fact that the branes are localized at the resolution ($\rho=a$),
and on the resolution we make the choice (without loss of generality) that they are at the North pole ($\xi=0$).
The equation to be solved takes the form
\bea
\nabla H\equiv\nabla_\rho H +{1\over \rho^2}\nabla_\xi H= -{k C \over 2\pi^4 \rho^7
\sin^5\xi\cos\xi } \delta(\rho-\rho_0)\delta(\xi)
\eea
over the resolved transverse space. We will set $k=4$ in the following as
we are discussing the $\IZ_4$ singularity.
The Laplace operators are
\bea\label{sm1}
&&
\nabla_\rho H\equiv
\big(1-{a^8\over \rho^8}\big)^{1\over 2}
\partial_\rho
\big(1-{a^8\over \rho^8}\big)^{1\over 2} \partial_\rho H+\frac{1}{\rho}\big(7-2\frac{a^8}{\rho^8}\big)\partial_\rho H, \br
&&
\nabla_\xi H\equiv \partial^2_\xi H + (5\cot \xi-\tan \xi) \partial_\xi H. \label{ang}
\eea
Let us define for  convenience $\sqrt{g_\rho}=\rho^7$ and $\sqrt{g_\xi}=
\sin^5\xi\cos\xi$.
In order to solve the full Green's equation with source, we first solve for the eigen-value
equation for the angular part $\nabla_\xi Y_l=-E_l Y_l$,
where $Y_l$ satisfy
\bea\label{sm2}
&&
\int_0^{\pi\over 2} Y^\ast_l(\xi)
Y_{l'}(\xi)\sqrt{g_\xi}d\xi=\delta_{ll'}, \label{ang-norm}\\
&&
\sum_{l} Y^\ast_l(\xi) Y_{l'}(\xi_0)={1\over\sqrt{g_\xi}}\delta(\xi-\xi_0).
\eea
This was the basis for the expansion in (\ref{Hgeneric}).
The angular harmonics are hypergeometric functions
\bea\label{sm3}
Y_l(\xi) \sim ~ _2F_1(-l,3+l,1, \cos^2\xi)
\eea
with the energy eigenvalues $E_l=2l(2l+6) \ge 0$. It turns out that these specific Hypergeometric functions can be rewritten as Jacobi polynomials of the form $P^{(0,2)}_l(-\cos
2\xi)$. The normalization of these have to be fixed by using the orthonormality of
Jacobi polynomials (see Appendix B) and (\ref{ang-norm}), and the result is
\bea
Y_l(\xi)=\sqrt{2(2l+3)}P^{(0,2)}_l(-\cos 2\xi).
\eea
With these  $Y_l$ solutions, the next step will be to solve for the radial
part
\bea\label{sm4}
\nabla_\rho H_l(\rho,\rho_0) -{E_l\over \rho^2}H_l(\rho,\rho_0)= -{ 2C \over \pi^4 \rho^7 }
\delta(\rho-\rho_0)
\eea
This can also be written as
\bea\label{sm5}
\Big(1-{a^8\over \rho^8}\Big)\partial_\rho^2 H_l+{1\over \rho}\Big(7+{a^8\over
\rho^8}\Big)\partial_\rho H_l  -{2l(2l+6)\over \rho^2}H_l= -{ 2C \over
\pi^4 \rho^7 } \delta(\rho-\rho_0).
\eea
Solving the homogeneous equation, we get two inequivalent solutions
\bea\label{sm6}
&& H_A \sim  \
{}_2F_1\left({3\over4}+{l\over4},-{l\over4};{3\over4};{\rho^8\over a^8}\right)\br
&& H_B \sim \ {\rho^2\over a^2} \
{}_2F_1\left({1}+{l\over4},{1-l\over4};{5\over4};{\rho^8\over a^8}\right)
\eea
These solutions have an exchange-symmetry under
$-l\leftrightarrow l+3$. The normalizations of the solutions are so far unfixed, and they need to be suitably chosen while
taking into account the delta function at the location of the M2-brane stack at $\rho=a$.

The first task in fixing the normalization is to find a linear combination of the two independent
solutions that dies down at infinity. By using certain Hypergeometric identities (see Appendix C), one can write a linear combination of $H_A$ and $H_B$ that manifestly has this property:
\bea
H_l(r)=\frac{C_l}{r^{2l+6}} {}_2 F_1 \left(\frac{l+3}{4},\frac{l+3}{4};\frac{2l+7}{4};-\frac{a^8}{r^8}  \right) \label{solve1}
\eea
Here $\rho^8=r^8+a^8$. To fix the overall normalization $C_l$, we need to integrate across the delta function. In the present case,
we need to solve
\bea
\left[\rho^7\Big(1-\frac{a^8}{\rho^8}\Big)\frac{{\rm d}H_l}{{\rm d} \rho}\right]\Big{|}_{\rho=a} \equiv \left[r(r^8+a^8)^{3/4}\frac{{\rm d}H_l}{{\rm d} r}\right]\Big{|}_{r=0}=-\frac{2C}{\pi^4}, \label{solve2}
\eea
to fix $C_l$. With the form for $H_l(r)$ from (\ref{solve1}), this can be solved explicitly in terms of Gamma functions:
\bea
C_l=\frac{C a^{2l}}{4 \pi^4}\frac{\Gamma\big(\frac{l+3}{4}\big)\Gamma\big(\frac{l+4}{4}\big)}{\Gamma\big(\frac{2l+7}{4}\big)}.
\eea

Using the fact that $ P^{(0,2)}_l(-1)=(-)^l(l+1)(l+2)/2$, we can finally write down the general solution for the M2 stack at the North pole as
\bea
H(r,\xi)=\sum_{l=0}^{\infty}(-)^l(l+1)(l+2)(2l+3)\ P^{(0,2)}_l(-\cos 2\xi) \ H_l(r).
\eea

\subsection{Membrane RG Flow}

We can check that this reduces to the smeared solution obtained before by restricting to the $l=0$ harmonic:
\bea
H(r,\xi)|_{(l=0)}=6 H_{l=0}=\frac{2C}{\pi^4r^6} \ {}_2 F_1\Big(\frac{3}{4}, \frac{3}{4};\frac{7}{4};-\frac{a^8}{r^8}\Big).
\eea
This looks superficially different from the smeared solution (\ref{smeared}), but in fact is the same as can be checked by expanding both expressions in a power series and comparing or by plotting them on Mathematica for various values of $a$. %{\bf (Check it against the smeared solution calculated earlier.)}.

It can also be checked that this singularity at $r=0$ arising from the
smearing is removed by the sum over the various $l$'s. For small values of $r$,  $H_l$ presented above is approximated by
\bea
H_l\sim -\frac{C \log(r)}{a^6},
\eea
and therefore, the full Green function takes the form
\begin{eqnarray}
H\sim -\frac{C}{a^6}\log(r)\sum_{l=0}^{\infty}2(-1)^l(l+2)(2l+3)
P_l^{(0,2)}(-\cos 2\xi)
\sim -\frac{C}{a^6}\log(r)\frac{\delta(\xi)}{\sqrt{g_\xi}}, \hspace{0.1in}
\end{eqnarray}
where in the last step, we have used the completeness relation (\ref{orth})
for Jacobi polynomials presented in the Appendix. In doing this, we are using $\xi_0=0$, and therefore  $ P^{(0,2)}_l(-\cos 2\xi_0)=(-)^l(l+1)(l+2)/2$. Two useful elementary delta function relations are $f(x)g(y) \delta(x-y)=f(x)g(x)\delta(x-y)$ and $\delta(f(x))=\frac{\delta(x)}{|f'(x)|}$.\footnote{In general, this last expression should take the form $\delta(f(x))=\sum_i\frac{\delta(x_i)}{|f'(x_i)|}$ where the sum is over the zeros of the function. But the relevant function in our case is $(1-\cos 2\xi)$ whose only zero in the range $[0, \pi/2]$ is at $\xi=0$. }

The above result makes it immediately clear that the
singularity in the smeared case at $r=0$ is removed because of the
vanishing of the delta-function away from the
location of the stack ($\xi=0$). The smearing of the
source branes on the six-cycle ($\IC\IP^3$) is also evident because the radial
part takes the form  $ \log r$, which is nothing
but the Green's function in (the remaining) two dimensions.

It turns out that by keeping track of the $l$-dependence of the
$H_l$ in the sum, we can extract a bit more information. The sum of all the various
$l$ pieces near $r=0$ should gives rise to an AdS
throat, because now we are around a smooth point. The
emergence of the throat is easy to
see if we approach $r=0$ along
$\xi=0$, because then the warp factor looks like
\eqn{wn}{H(r)=\sum_{l=0}^{\infty}\frac{(l+1)^2(l+2)^2(2l+3)}{2}H_l(r) \sim \sum_{l=0}^{\infty}l^5 H_l(r).}
We want to consider the near-horizon behavior where the local curvatures
are irrelevant, which means we are working in the limit where
the distance scales are much less than the resolution size, $r \ll a$. We
can solve the radial equation (away from the source) in this limit. In this limit
the homogeneous part of (\ref{sm1}) reduces to
\bea
\frac{a^8}{r^6}\frac{{\rm d}^2 H}{{\rm d} r^2}+\frac{a^8}{r^7}\frac{{\rm d}H}{{\rm d} r }-2l(2l+6)H=0 %-\frac{Ca^3}{2\pi^4r^7}\delta(r)
\eea
It turns out that the solution that dies down at infinity can be expressed in terms of
modified Bessel functions of the second kind. We will not present the explicit solution, except to note one crucial feature of the solution that will be important to us: the entire dependence of the
solution on $r$ and $l$ is captured by the combination
$\sqrt{l(l+3)}\ r^4 \sim l \ r^4$.  The normalization, which is fixed by integrating
the solution across the delta function as before, turns out to be independent of $l$. So we can write
\eqn{www1}{H(r) \sim \sum_l^{\infty} l^5 f(lr^4).}
Since we know that this sum has to be convergent in $l$, we can treat
the function $f(lr^4)$ as a regulator \cite{KM}. The way in which such a regulator accomplishes
finiteness is by decaying rapidly for $l > \frac{1}{r^4}$.  This means that we can approximate
the sum as
\eqn{www2}{H(r) \sim \sum_{l=0}^{1/r^4}l^5 f(lr^4) \sim \int_0^{1/r^4}
l^5 f(lr^4)dl\sim\frac{1}{r^{24}}\int_0^1 x^5f(x)dx = \frac{{\rm
const.}}{r^{24}}.}
We have approximated the sum by an integral and then done a change of variables. In the final step we have used the fact that
for the modified Bessel function mentioned earlier, the integral converges. (In fact for not too large $x$, the function can be approximated by
$\log x$.). Now all that we need to do in order to see the emergence of the throat, is to notice that close to $\rho=a$ (with
$\xi =0$), the metric (\ref{metricr}) takes the flat form with a
new radial coordinate $u\sim {r^4}$. Instead of systematically constructing the locally flat coordinates, one way to see this quickly is to expand the metric as
\bea
\label{smooth}
ds^2 \sim du^2+u^2d\beta^2+a^2ds^2_{\IC\IP^3}.
\eea
Since we are taking the near-horizon limit along $\xi=0$ (see the comment before \ref{wn}), the last piece can safely be dropped and the radial coordinate effectively becomes $u$. \footnote{Another way to say the same thing is to note that around a smooth point in (\ref{smooth}) a flat space radial coordinate can be defined by $\sqrt{u^2+a^2 \xi^2}$. This is because the radial coordinate of a space with metric of the form $ds^2= dr_1^2+r_1^2d\Omega_1^2+dr_2^2+r_2^2d\Omega_2^2$ is $\sqrt{r_1^2+r_2^2}$, as can be easily verified by translating to Cartesian coordinates.} So in terms of this flat
coordinate, the
warp factor (\ref{www2}) goes as $\sim \frac{1}{u^6}$. But a hexic fall-off is precisely what is needed to generate the $AdS_4$ throat, cf., equations (\ref{warp1}, \ref{S5z3}). So finally, we end up with $AdS_4\times S^7$ as expected, in the near-horizon limit around a smooth point.

\section{Chern-Simons Level Shift}

By gauge-gravity duality, we expect that the supergravity solution we constructed in the previous section should have an interpretation in the field theory. The dual gauge theory for M2-branes on a $\IC^4/\IZ_k$ orbifold is given by ABJM theory at a Chern-Simons level $k$. Turning on a VEV corresponds to moving the branes away from the singularity. When we do this, far away from the resolution, the gauge theory should still look like the dual of the theory on the $\IZ_k$ orbifold, but if we zoom in on the stack, we expect that the dual theory should have $k=1$, because the membranes are effectively in flat space. The explicit supergravity solution we constructed connecting these two cases should correspond, on the gauge theory side, to an RG flow. Our purpose in this section is to see how this expectation is realized, by finding pieces of evidence for the RG flow triggered by the VEV and to see the reduction in the Chern-Simons level. %We will find that the RG flow that is realized has some interesting differences compared to the AdS$_5$/CFT$_4$ case. %We also emphasize that much of the material we discuss is well-known, our aim is to put them in a context that is of interest to us.
We will first consider the case of generic $k$ with the gauge group rank $N=1$, in which case we do not have to deal with the complications arising from the ABJM superpotential because it vanishes identically. %\footnote{We will make some comments about the more general case later. }.
Notice that the supergravity limit corresponds to the large $N$ limit (for fixed small $k$), so the situation in this section captures a different regime. %But we also mention that the analysis of moduli spaces in the $U(1)$ cases in AdS$_5$/CFT$_4$ usually captures most of the interesting information. For the moment, we will work with general level $k$. In the next subsection we will specialize to the $k=4$ and consider RG flows on the resolved orbifold.
The emergence of the moduli space in ABJM has been considered in \cite{ABJM}, generalizing \cite{Tong}. %by redefining the gauge fields according to \cite{Benna}:\bea B_\mu\equiv \hat A_\mu -A_\mu, \ \ {\cal A}_\mu \equiv\hat A_\mu + A_\mu. \eea With these definitions, the bosonic part of the ABJM action from section 2 (in the Abelian case) becomes: \bea \label{action} {\cal L}=-D_\mu \bar C^I \bar D^{\mu}C_I - \frac{k}{4 \pi} \epsilon^{\mu\nu\rho}B_\mu F_{\nu\rho}. \eea Here $F_{\mu\nu}$ is the field strength of ${\cal A}_\mu$, the scalar fields are written in the $SU(4)$ notation, and the covariant derivative now turns into $D_\mu=\partial_\mu -iB_\mu$, and it depends only on $B$. The potential term identically vanishes in the $U(1)$ case. We will write this Lagrangian as that of a sigma model so that we can read off the moduli space directly. In order to do this, we can do some field redefinitions. First, we will treat the field strength $F_{\mu\nu}$ as a fundamental field. But in order to do so, we need to make sure that the Bianchi identity is imposed as a constraint. We do this through a Lagrange multiplier field and add to (\ref{action}) a piece \bea \label{newact} {\cal L}=-D_\mu \bar C^I \bar D^{\mu}C_I - \frac{k}{4 \pi} \epsilon^{\mu\nu\rho}B_\mu F_{\nu\rho}-\frac{\tau (x)}{4\pi}\epsilon^{\mu\nu\rho}\partial_\mu F_{\nu\rho}. \eea Here $\tau$ is the Lagrange multiplier and it is $2\pi$-periodic. This can be seen as follows. If we shift $\tau$ by a constant $\alpha$, then the action changes by \bea \delta S=\frac{\alpha}{2\pi}\int d^3 x \epsilon^{\mu\nu\rho}\partial_\mu F_{\nu\rho}=\frac{\alpha}{4\pi} \int_{\partial M} F= \alpha \ m, \eea where $m \in \IZ$. In the second equality, we have used Stokes' theorem and in the final step, the fact that the flux is quantized. So if $\alpha=2\pi$, the action changes by $2\pi$ and therefore path integral remains unchanged. This shows that shifting $\tau$ by $2\pi$ is a redundancy of the theory. This will be useful to us momentarily. After an integration by parts, the Lagrangian (\ref{newact}) takes the form \bea {\cal L}=-(\partial_\mu-iB_\mu)\bar C^{I}(\partial^\mu+iB^\mu)C_I -\frac{1}{4\pi}(k B_\mu+ \partial_\mu \tau) F_{\nu \rho}. \eea The equation of motion for $F$ now forces the relation \bea B_\mu=-\frac{1}{k}\partial_\mu \tau. \eea Using this, we can rewrite the action in terms of the new field $ Y^I \equiv e^{i \tau/k}C^I$ in the very simple final form \bea {\cal L}=-\partial_\mu \bar { Y}^I \partial^\mu { Y_I}. \label{ABJMnew} \eea Since $I$ runs from 1 to 4, at this point the moduli space looks like $\IC^4$. But we haven't fully taken care of the gauge invariance yet. We note that the gauge transformation property \bea \delta B_\mu=\partial_\mu \Lambda \ \ {\rm implies \ that} \ \ \delta \tau = -k \Lambda. \eea Therefore, under the gauge transformation, since $C^I \rightarrow e^{-i\Lambda}C^I$, the field $ Y^I$ remains invariant. But the $2\pi$-periodicity of $\tau$ implies that  $ Y^I \sim e^{\frac{2\pi i }{k}} Y^I$. This means that the target space is the $\IC^4/\IZ_k$ orbifold and that $U(1)$ ABJM theory (ignoring fermions) is precisely the sigma model on this space.
It turns out from their work that we can write the action for level $k$ ABJM theory for $N=1$, as a sigma model
\bea {\cal L}=-\partial_\mu \bar { Y}^I \partial^\mu { Y_I}.\label{ABJMnew} \eea The fact the moduli space is $\IC^4/\IZ_k$ is captured in this language by a redundancy in $Y$: one should identify $Y^{I} \sim  e^{\frac{2\pi i }{k}} Y^I$.
Our intention is to see how the theory changes when we give a vev that corresponds to moving the branes away from the orbifold point. To see what happens, it is easiest to write the sigma model in a coordinate system where this motion is easily implemented. A convenient coordinate system is the one presented in (\ref{compcord}-\ref{compcord4}). In terms of these new fields, the action takes the form $
{\cal L} \sim (\partial r)^2 +r^2 (\partial \Omega_{S^7/\IZ_k})^2$,
where $\partial \Omega_{S^7/\IZ_k}$ is a shorthand for the various pieces in the angular part\footnote{The angular part is that of the orbifold, but this does not affect any of the arguments below because the new fields get appropriately rescaled.}. Notice that $r^2 = \sum |Y^I|^2=\sum |C^I|^2$, so moving away from the orbifold corresponds to giving a vev to (at least) one of the fields as expected. Now it is easy to see that the when we give a vev as $r=v+r'$, the action can be rewritten as that of flat eight dimensional space, plus terms that are suppressed by higher powers of $\frac{1}{v}$. \footnote{This is just the familiar statement that all spaces look flat locally. A simple example that captures the basic idea is to consider the action $\partial \bar z \partial z$, for a complex field $z$. We can re-express it as $z=\rho e^{i\theta}$, and if we give a vev as $\rho=v+\rho'$, then the action takes the form $(\partial \rho')^2+(\partial \theta')^2+{\cal O}(\frac{1}{v})$, where $\theta'\equiv v \theta$ in order to retain canonical kinetic terms. So the new coordinates $\rho'$ and $\theta'$ are flat 2D Cartesian coordinates, despite their names. Our situation is exactly analogous, except we have more angular coordinates.} The punch-line then, is that after giving the vev corresponding to the motion away from the tip, the theory reduces to that of a sigma model on flat eight dimensional space (up to higher order terms which are suppressed in the IR). But such a theory is nothing but setting $k=1$ in the $U(1)$ ABJM theory as defined by (\ref{ABJMnew}). Thus starting from the generic $k$ Abelian ABJM theory, and by giving the appropriate vev, we have flown to ABJM at level 1. This is precisely what the supergravity solution we constructed captured in a different regime.

The discussion above was for the Abelian case. For the generic case of $U(N)\times U(N)$, the ABJM superpotential does not vanish, and the Abelian duality we did earlier in order to write down the sigma model variables does not go through. %The emergence of the full ABJM theory with a shifted level is difficult to demonstrate.
Because of the non-trivial superpotential, the translation in the non-Abelian case is not clear to us. %A related issue is the fact that the quantity $TrC^I$ which one gives vev to in order to get the $M2\to D2$ transition, is not a gauge invariant quantity, unlike in the SYM case. It is charged under the off diagonal part of $U(N)\times U(N)$. In order to properly move on the single membrane moduli space, one should correctly identify the center of mass variable and give vev to it. In the $U(1)\times U(1)$ case the center of mass is given by the gauge invariant ``dressed'' field $Y=e^{i \tau/k}C$ which is non-locally related to the original degrees of freedom. (This is because $\tau (x) \sim\int^x dy^\mu B_\mu=\int^x dy^\mu (A_\mu-\hat A_\mu)$).
It is known that the chiral operators of ABJM theory contain Wilson lines \cite{ABJM}, and therefore the problem might involve a non-local redefinition of fields: the analog of the $Y$ operator might contain Wilson lines  connecting  $C(x)$ to $\infty$. In the Abelian case, the ``dressed'' field in the sigma model variables $Y$ that we wrote down, contains $\tau (x) \equiv % \int^x dy^\mu B_\mu=
\int^x dy^\mu (A_\mu-\hat A_\mu)$ in terms of the original ABJM variables \cite{ABJM}. Supersymmetric Wilson lines in ABJM have been considered in \cite{Drukker, Chen, Rey, Berenstein, Kluson}. See \cite{Kleb2} for a related discussion in terms of 't Hooft operators. %it seems likely that one might be able to cook up a dressed field by inserting (supersymmetric) Wilson lines such that the Laplacian of this new operator vanishes as a result of the ABJM equations of motion. This will be the free center-of-mass mode that we are looking for.
But the full gauge-invariance and  path-independence of such an operator are non-trivial and it is not clear to us if one can isolate its dynamics from the rest of the fields. Some of these questions are currently being investigated.

Despite these difficulties, there is one simple check that we {\em can} do even in the non-Abelian case: we can check that the moduli space of the theory that results after we turn on the vev, reproduces the expected moduli space at $k=1$. The moduli space of the theory before we give the vev is $(\IC^4/\IZ_k)^N/S_N$. This corresponds to the matrices $C^I$ being diagonal. This breaks the gauge symmetry to $U(1)^N \times U(1)^N$ times the permutation of the diagonal elements. The off-diagonal elements all get mass, which means that only the $U(1)^N \times U(1)^N$ part of the $U(N) \times U(N)$ theory is relevant to the discussion of the moduli space. Since the $U(1)^N \times U(1)^N$ part is nothing but $N$ copies of the $U(1)\times U(1)$ theory with the same flux quantization conditions as in the Abelian version \cite{ABJM}, we can write it as $N$ copies of (\ref{ABJMnew}). So to see the new moduli space for a theory expanded around the generic vev, we can reuse the arguments for the Abelian case, which gives us now $N$ copies of $\IC^4$ up to permutations. Therefore we see that the new moduli space is $(\IC^4)^N/S_N$, which is the moduli space of $N$ M2-branes in flat space.

The situation we have considered in this section ($v \rightarrow \infty$), is very different from the case when we give a vev $v \rightarrow \infty, k \rightarrow \infty$, while holding $v^2/k$ fixed. Large $k$ effectively reduces the transverse size of the cone (or more precisely the $U(1)$ fibration), and if we consider this as the M-theory circle, then we are effectively left with a theory of D2-branes \cite{Mukhi}, with Yang-Mills coupling $\sim v^2/k$. In the finite $k$ case which we are considering, if we naively give a vev like in \cite{Mukhi}, but without tuning $k$, the theory still turns into ${\cal N}=8$ SYM plus more IR suppressed terms. What we want therefore is the IR limit of ${\cal N}=8$ SYM, which cannot be captured by a purely classical analysis. 
%What we want is the IR limit of this theory, which we expect is the correct ABJM theory. 
It should also be noted that usually Chern-Simons level shifts refer to a one-loop shift. See, e.g., \cite{Berman} for a discussion in the context of BLG theory. The level shift that we consider here arises after field re-identifications along the RG flow which starts at generic $k$ and ends at $k=1$.

It is instructive to compare our arguments above with the situation in $AdS_5/CFT_4$. There, the moduli space is usually expressed in terms of independent gauge invariant polynomials constructed out of the chiral bi-fundamentals. Of all such operators constructed that way, many are redundant, because of the F-term conditions. The algebra of the rest fixes the moduli space of the theory\footnote{The construction of the moduli space is analogous to the algebraic construction of, e.g., the space $\IC^4/\IZ_4$, presented in (\ref{monom}) in appendix D. But note that in ABJM, the algebraic construction of the space is quite different from the construction of the moduli space in the gauge theory, as done in \cite{Tong, ABJM}. }. Giving a vev to one of the fields triggers an RG flow that finally leads one to ${\cal N}=4$ SYM. But in ABJM theory, the bi-fundamentals directly define a sigma model, and the vev changes the level of the theory.

\subsection{The Gauge Dual of the Resolved Orbifold}

In the case of the conifold, the Klebanov-Witten gauge theory \cite{KW1} is supposed to be dual to the unresolved conifold. In fact, moving the D3-branes away from the tip of the conifold corresponds to giving a vev to the bifundamental scalars of the theory ($A_i, B_i$), but in such a way that the operator
\bea
\label{KWD}
\frac{1}{N}{\rm Tr}(|A_1|^2+|A_2|^2-|B_1|^2-|B_2|^2)
\eea
remains zero. If this operator is not zero, then the gravity dual was interpreted to be no longer the singular conifold, but the resolved conifold \cite{KW2}. Can we say something analogous in our case?

We can try to identify the dual gauge theory operator that is being turned on (when we resolve the orbifold) by looking at the linearized solution around the AdS background.
The AdS/CFT dictionary relates the asymptotic fall off of the bulk fields to the dimensions of the gauge theory operators. Since we are interested only in the asymptotics, we can restrict our attention to the smeared form of the warp factor (\ref{smeared}). From the explicit expression, we find that
\bea
H\approx {L^6 \over \rho^6}\left(1+ {3a^8 \over 7\rho^8} + ...\right),
\eea
where dots indicate higher order terms. Using this in the standard M2-brane ansatz
(\ref{m2ansatz}), using (\ref{metricr}) for the transverse metric, we end up with
\bea
ds^2\approx{\rho^4 \over L^4} \eta_{\mu\nu}dx^{\mu}dx^{\nu}+\frac{L^2}{\rho^2}d\rho^2+L^2ds^2_{S^7/\IZ_4} +\hspace{0.55in}\nonumber \\
-{2a^8 \over 7L^4\rho^4}\eta_{\mu\nu}dx^{\mu}dx^{\nu}+\frac{8L^2a^8}{7\rho^{10}}d\rho^2+
\frac{8L^2a^8}{7\rho^{^8}}(ds^2_{S^7/\IZ_4}+...)%+\frac{L^2a^8}{\rho^{10}}
%-{L^2a^8 \over \rho^{10}}\left({d\beta - A\over 4}\right)^2
+...
\eea
The first line reproduces precisely the $AdS_4\times S^7/\IZ_4$. And it is clear from comparing with this piece that the fall-offs in the second line correspond to that of a dimension 4 operator in the dual gauge theory\footnote{The AdS-CFT correspondence, as presented in \cite{adscft}, dictates that the dimension of the operator is fixed by the fall-offs through the coordinate $z \sim \frac{1}{\rho^2}$ defined before (\ref{S5z3}).}. Note also that the relative normalization is what determines the dimension. This is quite analogous to the case of the blown up four-cycle considered in \cite{benvenuti}. Our resolution here is a six-cycle, as can be seen purely algebraically, from the discussion given in Appendix D. It is easy to check also that the fall-off of the form field $C_{012} \sim H^{-1}$ also leads to the same result for the dimension.
%the gauge-gravity theory of blown-up four cycles in $AdS_5/CFT_4$ has been investigated in \cite{benvenuti}.

In the above, we have worked with the smeared solution. But it is possible to argue directly using the K\"ahler potential (\ref{kahler}) that the answer is unchanged even without this assumption. This is analogous to the discussion around eqn. (4.16) in \cite{KW2}. What we need to look at is the expansion of (\ref{kahler}) to lowest order in $a$. For the case $n=4$ that is relevant here, we immediately see from the properties of Hypergeometric functions that
\bea
K(r) \sim r^2 \Big(1 - \frac{1}{12}\frac{a^8}{r^8}+ {\cal O}(r^{-16})\Big)
\eea
So we see that the first corrections arise at order ${\cal O} (r^{-8})$, i.e., dimension 4. When we compute the metric using the K\"ahler potential, the angular parts can at most {\em increase} the rate of fall-off if they have the correct symmetry properties and sum up to zero at some order, but never reduce it. In particular, the smeared solution captures the averaged out fall-off over the angles, and since it starts at ${\cal O} (r^{-8})$, we again come to the conclusion that the dimension of the dual gauge theory operator is 4. Note also that in the case of the resolved conifold, the smeared solution constructed by Pando-Zayas and Tseytlin \cite{tseytlin} had the same fall-off as the un-smeared version suggested in \cite{KW2} and explicitly constructed in \cite{KM}.

Can we be any more specific about the nature of this operator? In the case of the resolved conifold, it was possible to identify the operator because it was in a baryonic current multiplet, and therefore was protected. %In our case the baryonic symmetry is gauged because the gauge group is $U(N) \times U(N)$ and not $SU(N) \times SU(N)$.
The resolution parameter in Klebanov-Witten was related to a D-term in the gauge theory (see \ref{KWD}), and it had precisely the correct dimension, namely 2. In our case, the baryonic current\footnote{Note that the baryonic symmetry is gauged in ABJM, because the gauge group is $U(N) \times U(N)$ and not $SU(N) \times SU(N)$. } is of the form ${\rm Tr}(C^{I\dagger}C^I)$, and has dimension 1 (we are in 3D). This is clearly not the dimension of the operator dual to the resolution that we found above, suggesting that the baryonic current is not the correct operator. This is not surprising. In general, the dimensions of the operators dual to resolutions are unknown even when there are baryonic symmetries, see \cite{benvenuti} for a discussion of general four-cycles in $AdS_5/CFT_4$.

Another way to motivate that the construction in ABJM has serious differences from that in KW is as follows. In the analogue of ``D-terms" in three dimensions, the role of the FI parameter is captured by the scalar in the vector multiplet \cite{cft3}. %, which is a {\em field}.
This is one manifestation of the fact the  determination of the moduli space in ABJM theory crucially depends on the gauge redundancy. It should also be noted that once we take these gauge redundancies correctly into consideration, the moduli space is no longer the conifold, but an orbifold, despite the fact that the superpotentials have the same form \cite{ABJM}. In KW on the other hand, the F-term relations arising from the superpotential immediately determine the moduli space to be the conifold. In ABJM case (with $k=4$), the moduli space arises from the complex coordinates through the identification $Y^I \sim e^{i\frac{\pi}{2}}Y^I$ arising from the gauge-redundancy \cite{ABJM, Tong}. The invariant monomials that one can construct for this orbifold action satisfy the complicated algebra of the $\IC^4/\IZ_4$ orbifold\footnote{See appendix D for the monomials, their algebra is straightforward enough to write down, but we don't write them explicitly because they are cumbersome and not immediately useful.}, and not the rather simple expression for the conifold (namely, $xy=zw$), and therefore a characterization of the resolution will be substantially more complicated. With only such an implicit description, the determination of the operator dual to the resolution is a non-trivial task, even in the off-chance that it falls in a protected multiplet of some sort. We will not pursue this direction further here, but merely remark that it is certainly worthy of further study.
%it should be noted that the algebraic construction of the space considered in

Finally, there is the problem that our resolution does not preserve as much supersymmetry as the ABJM orbifold does. It would be interesting to construct a resolution of the ABJM orbifold in the natural complex structure ( $\{ w_a, \bar w_b \}$ instead of $\{w_a, w_b\}$) where perhaps one can have more control on the problem both from gauge theory and gravity. 

Another related comment is that baryonic operators in ABJM are more subtle than in the conifold theory. An operator of the schematic form $\det C$, which is the analogue of the conifold baryonic operator, is not gauge invariant because the gauge group is $U(N) \times U(N)$ and not $SU(N) \times SU(N)$. It has been suggested that baryonic operators in ABJM theory can be constructed by the insertion of Wilson lines in $\det C$ \cite{Park}. Aspects of baryonic operators and symmetries in $AdS_5/CFT_4$ have been investigated in \cite{Gukov}.

Our orbifold is toric, and so in algebraic language, the resolution can be implemented in terms of the charge zero monomials that are invariant under the orbifold action. Some of the details are presented in an appendix. In the toric language the Higgsing in the resolved case corresponds to zooming in on a specific patch, say $z_4 \ne 0$ on the blown up $\IC\IP^3$. We can define new coordinates on this patch according to\footnote{All the 35 monomials listed in the appendix can be re-expressed as monomials of these new variables satisfying the same algebra, as can be easily checked.}
\bea
u_1=\frac{z_1}{z_4}, \ u_2=\frac{z_2}{z_4}, \ u_3=\frac{z_3}{z_4}, \ u_4={z_4^4z_5}.
\eea
Notice that this splits off the $\IC\IP^3$ (the first three coordinates are nothing but the coordinates on the standard chart on the $z_4 \ne 0$ patch of $\IC\IP^3$) and the forth coordinate is a $\IC$ fiber on the $\IC\IP^3$, with an appropriate transition function piece that captures the Chern class of the bundle. In fact, this is the total space of the line bundle ${\cal O}(-4)\rightarrow \IC\IP^3$. When we set $|z_4|\sim \mu$, we only see the $(u_1,...,u_4)$ patch, which is nothing but $\IC^4$. In the toric diagram in figure 1 (see Appendix D), letting $\mu \rightarrow \infty$ is equivalent to chopping off the node $[-1,-1,-1,1]$ and we end up with the toric polytope for $\IC^4$.

Since we do not know the precise form of the operator dual to the resolution, we cannot be sure what vevs to give in order to trigger the RG flow. But we can try to see whether giving a vev analogous to the one we tried in the unresolved case is a reasonable choice. Indeed, if we give a ``vev" \footnote{Note that $Y^I$ is not gauge-invariant, and this is merely a schematic argument. This should be compared to the conifold, where the gauge-invariant operators are constructed from traces of products of fields, but it is sometimes useful to talk about giving a ``vev" to $A$ or $B$, in such a way that gauge-invariant operators end up getting non-zero vevs.} to one of the $Y^I$, it breaks the $SU(4)$ down to $SU(3)$ and this is preserved by a localized stack on the resolution, because $\IC\IP^3$ is the coset space $SU(4)/U(3)$. (This is analogous to the fact that if we put a localized source on the sphere $S^2$, there is an unbroken $SO(2)$ that is left of the original $SO(3)$: this is because $S^2=SO(3)/SO(2)$). In this simple case, the natural candidate for the endpoint of the RG flow is ABJM at $k=1$, but it would be useful to have a more hands-on understanding of the path of these flows. In particular, in more general cases, it would be interesting to understand whether the RG flows in M2-brane theories demonstrate structures (cascades, walls, trees, ...) analogous to the RG flows in four dimensional $N=1$ theories.

\section{Summary and Comments}

%Since we discussed the conclusions in the introduction, we make some comments in this section regarding directions for future work.

In this paper we have looked at M2-brane stacks, on certain orbifolds and their resolutions. Our interest was primarily in seeing the RG flows that are created when we move the stack to a smooth point. On the supergravity side, we have constructed explicit solutions that capture this RG flow and demonstrated that they indeed interpolate between the appropriate limits. Our discussion of the gauge theory side was far less exhaustive, but we have presented some arguments why it is reasonable for the Chern-Simons level of ABJM gauge theory to undergo a shift due to the RG flow. We have also constructed supergravity solutions dual to brane stacks on a resolved orbifold. This corresponds to a dimension 4 operator in the dual gauge theory. The emergence of the $AdS_4 \times S^7$ throat is analogous to similar constructions in $AdS_5/CFT_4$. But unlike in $AdS_5/CFT_4$, where the end result of such a flow is ${\cal N}=4$ SYM,   here the RG flow results in a shift to $k=1$ in the Chern-Simons level, at the IR fixed point.

In the rest of the section, we make some comments about possible directions for future work.

In the part of this paper where we considered the resolution, we focussed on the orbifold $\IC^4/\IZ_k$ for the specific case $k=4$. The reason was that for this value of $k$, the metric on the orbifold (and its resolution) can easily be found generalizing the Eguchi-Hanson metric on $\IC^2/\IZ_2$ in four dimensions (as we sketched in a previous section). It would be very interesting to construct explicit metrics for the resolutions when $k$ is generic. The metric on the unresolved space $\IC^4/\IZ_k$ was written down in \cite{hyperkahler, ABJM} using the fact that it can be thought of as a toric hyperK\"ahler manifold. These spaces are related to Kaluza-Klein monopoles that generalize Gibbons-Hawking (Taub-NUT) spaces in four dimensions. Gibbons-Hawking metrics have an orbifold singularity when the many centers coincide, which gets resolved when the centers are moved apart. So it seems plausible that one could construct resolutions of $\IC^4/\IZ_k$ by manipulating the various ``centers" that arise in its construction as a  toric hyperK\"ahler manifold.

We have tried to explore some aspects of the RG flow in the gauge theory side. We have investigated only the Abelian case (i.e., the case when the gauge group is $U(1) \times U(1)$) where the superpotential identically vanishes and an Abelian dualization on the gauge-fields can be usefully executed to understand what happens when we turn on a VEV. In the case when the gauge group rank is higher, the superpotential is non-trivial, and the situation is more complicated. Since the chiral operators of the theory contain a Wilson line that goes to infinity, finding a manifestly local description is probably difficult. It would be interesting to understand these questions better.

\section*{Acknowledgements}

We would like to thank Jarah Evslin, Amihay Hanany, Stanislav Kuperstein and K. Narayan for discussions. This work is supported in part by IISN - Belgium (convention 4.4505.86), by the Belgian National Lottery, by the
European Commission FP6 RTN programme MRTN-CT-2004-005104 in which CK and CM are associated with V. U. Brussel, and by the Belgian Federal
Science Policy Office through the Interuniversity Attraction Pole P5/27.

\section{Appendix}
\subsection*{{\bf A.} \ $\IC\IP^3$ and $S^7$: Factsheet}
\addcontentsline{toc}{subsection}{{\bf A} \ $\IC\IP^3$ and $S^7$: Factsheet}
\renewcommand{\theequation}{A.\arabic{equation}}

The standard Fubini-Study metric on $\IC\IP^3$ can be written in terms of
the
three complex projective coordinates $\zeta_i$: %$\equiv \omega^i/\omega_4$
 \begin{equation}
ds_{\IC\IP^3}^2={ d\bar\zeta_i d\zeta^i\over \rho^{2}} -
{(\zeta^id\bar\zeta_i)
(\bar\zeta^j  d\zeta_j)\over \rho^4}
\end{equation}
where $\rho^2=1+\bar\zeta^i\zeta_i$.  %Thus $\IC\IP^3$ is described as a coset space $SU(4)/U(3)$.
In terms of the parametrization for $\IC^4$ that we introduced in (\ref{compcord})-(\ref{compcord4}), we can define explicit coordinates $\zeta_i$ on $\IC\IP^3$ by going to a specific patch. If we choose
a patch where $w_4$ in (\ref{compcord}) is non-vanishing, then we can define
\bea
\zeta_i=\frac{w_i}{w_4} \ \ {\rm for} \ \ {i=1,2,3}
\eea
and the metric takes the explicit form
\begin{equation}
ds_{\IC\IP^3}^2=d\xi^2 +\sin^2\xi\left( d\alpha^2 +{1\over 4}
\sin^2\alpha(\sigma_1^2+\sigma_2^2+\cos^2\alpha\,\sigma_3^2)+{1\over4}
\cos^2\xi(d\chi+\sin^2\alpha\,\sigma_3)^2\right)
\end{equation}
The $\sigma_i$ are the left-invariant one-forms of $SU(2)$ (There is an
$S^3$ inside $\IC\IP^3$ and it is useful to think of it as the group
manifold $SU(2)$ for parametrization purposes):
\bea
\sigma_1 &=& \cos \psi\ d\theta + \sin \psi\ \sin \theta\ d\varphi ,\\
\sigma_2 &=& - \sin \psi\ d\theta + \cos \psi\ \sin \theta\ d\varphi ,\\
\sigma_3 &=& d\psi + \cos \theta\ d\varphi .
\eea
They are chosen so that they satisfy the Maurer-Cartan equation for the
group:
\bea
d\sigma_i = -\frac{1}{2}\epsilon_{ijk}\sigma_j \wedge \sigma_k
\eea
%where $\sigma_i$'s are the three left invariant 1-forms satisfying the $SU(2)$ algebra.
The ranges/periodicties of the angles are $0\le \xi, \alpha \le \frac{\pi}{2}, 0\le \theta \le \pi, 0\le \varphi \le 2\pi$ and $ 0\le \psi, \chi \le 4\pi$. The metric above is identical to the one in, for example, \cite{Pope}. Some useful references are \cite{Cvetic, Lukas}.

We can now introduce a round seven-sphere as a fibered $U(1)$ bundle over
$\IC\IP^3$ base
\begin{equation}
ds_{S^7}^2= (d\beta' -A')^2 + ds_{\IC\IP^3}^2
\end{equation}
where $0\le\beta'\le2\pi$ is the range of the coordinate on the fibers. %and the K\"ahler 2-form is locally given by
The form $A'$ is defined in terms of the K\"ahler form of the base and can be taken in the form \cite{Pope} $
A'={1\over2} \sin^2\xi(d\chi+\sin^2\alpha\sigma_3)$.
It is now easy to define a metric on $S^7/Z_k$ as
\begin{equation}
ds_{S^7/Z_k}^2= {1\over k^2}(d\beta - A)^2 + ds_{\IC\IP^3}^2
\end{equation}
where $0\le\beta\le2\pi$ and
\begin{equation}\label{k=4}
A={k\over2} \sin^2\xi(d\chi+\sin^2\alpha\sigma_3)
\end{equation}

The invariant volume element on $\IC\IP^3$ can be written as
\bea
\sqrt{g_{\IC\IP^3}}= {1\over 16}
\sin^5\xi\cos\xi\sin^3\alpha\cos\alpha\sin\theta.
\eea
The volume of unit $\IC\IP^3$ is then ${\pi^3\over 6}$ and that of
unit radius $S^7/Z_k$ is
${\pi^4\over 3 k}$.

%{\bf The relevant details of $\IC\IP^3$ (and its fibered 7-sphere) that I have used, I will add later. Everything we need, in case you want to take a look, is in \cite{Pope}.}

 It will sometimes be useful to have various geometrical results regarding $S^7$ with the metric induced from the Euclidean space into which it is embedded. This embedding done in the standard way according to the relation $\sum_{i=1}^{8}{(x^i)}^2 =r^2$. The metric of flat 8D space can then be written as $ds^2=dr^2+r^2d\Omega_{S^7}^2$, where $d\Omega_{S^7}^2$ is the metric on the unit 7-sphere. To do explicit computations, sometimes it is useful to work in spherical polar coordinates which are defined in the standard way as
\bea
x^1=r \cos \theta_1, \ \ x^2=r \sin \theta_1 \cos \theta_2, \ ... \ , \  %\ \ x^3=r \sin \theta_1 \sin \theta_2 \cos \theta_3,  \hspace{1in}\nonumber\\
%x^4=r \sin \theta_1 \sin \theta_2 \sin \theta_3 \cos \theta_4, \ \ x^5=r \sin \theta_1 \sin \theta_2 \sin \theta_3 \sin \theta_4 \cos \theta_5, \hspace{0.8in}\nonumber \\
%x^6=r \sin \theta_1 \sin \theta_2 \sin \theta_3 \sin \theta_4 \sin \theta_5 \cos \theta_6, \
x^7=r \sin \theta_1 \sin \theta_2 \sin \theta_3 \sin \theta_4 \sin \theta_5 \sin \theta_6 \cos \phi, \nonumber \\ {\rm and \ \ finally \ \ }
x^8=r \sin \theta_1 \sin \theta_2 \sin \theta_3 \sin \theta_4 \sin \theta_5 \sin \theta_6 \sin \phi. \hspace{1in} \nonumber %\hspace{1.7in}\nonumber
\eea
The coordinates $\{ r, \theta_1, ... , \theta_6, \phi \}$ span $\IR^8\equiv\IC^4$, when $r \in [0, \infty), \theta_i \in [0,\pi]$ and $\phi \in [0,2\pi)$. In these coordinates the metric on $S^7$ takes the explicit form
\bea
d\Omega_{S^7}^2
=d\theta_1^2+\sin^2\theta_1 (d\theta_2^2+ \sin^2\theta_2 (d\theta_3^2+\sin^2\theta_3 ( d\theta_4^2+ \sin^2 \theta_4 (d\theta_5^2 +  \sin^2 \theta_5 (d\theta_6^2 + \sin^2 \theta_6 d\phi^2))))). \nonumber
\eea
This leads to the 8-dimensional volume form $r^7 \sin^6\theta_1 ... \sin\theta_6 dr d\theta_1 ... d\theta_6 d\phi$. Integrating the angular part over their range, we find that the volume of the 7-sphere is ${\pi^4 \over 3}$. Notice that by volume here, we mean the volume of $S^7$ as a manifold %The high-school definition of ``volume" usually refers to the solid-sphere.
and also that it agrees with what was found through the $U(1)$ fibration over $\IC\IP^3$, as it should.

%\section*{Appendix: M2-branes on the Unresolved Orbifold, away from the Singularity}

\subsection*{{\bf B.} \ Jacobi Polynomials}
\addcontentsline{toc}{subsection}{{\bf B} \ Jacobi Polynomials}
\renewcommand{\theequation}{B.\arabic{equation}}

We collect some useful features of Jacobi polynomials (a kind of orthogonal polynomials) here. They show up in the angular part of the harmonics on $\IC\IP^3$.

For our purposes, Jacobi polynomials are defined in terms of Hypergeometric functions as
\bea
P_n^{(\alpha,\beta)}(x)=\frac{\Gamma(n+\alpha)}{\Gamma(\alpha)\Gamma(n)}{}_2F_1 \left(-n,n+\alpha+\beta+1;\alpha+1;\frac{1-x}{2}\right)
\eea
where $\Gamma(x)$ is the Euler Gamma function. For each choice of the pair of indices $\alpha, \beta$, we get an orthonormal set of basis functions. Their orthogonality relation takes the form
\bea
\int_{-1}^{+1}(1-x)^{\alpha}(1+x)^{\beta} P_m^{(\alpha,\beta)}(x)P_n^{(\alpha,\beta)}(x)\ {\rm d}x= \hspace{1.3in}\\
\nonumber \hspace{1.3in}= \frac{2^{\alpha+\beta+1}}{2n+\alpha+\beta+1}\frac{\Gamma(n+\alpha+1)\Gamma(n+\beta+1)}{n!\ \Gamma(n+\alpha+\beta+1)}\delta_{nm}.
\eea
The completeness relation is
\bea\label{orth}
\sum_{n=0}^{\infty}\frac{n! \ (2n+\alpha+\beta+1)\Gamma(n+\alpha+\beta+1)}{\Gamma(n+\alpha+1)\Gamma(n+\beta+1)}P_n^{(\alpha,\beta)}(x)P_n^{(\alpha,\beta)}(y)=  \hspace{1.3in}\\ \nonumber \hspace{1.3in}= 2^{\alpha+\beta+1}(1-x)^{-\alpha/2}(1+x)^{-\beta/2}(1-y)^{-\alpha/2}(1+y)^{-\beta/2}\delta(x-y).
\eea
In all the above relations, $\rm{Re} (\alpha, \beta) > -1$, and $n$ is a positive integer.

\subsection*{{\bf C.} \ Gamma Functions and Hypergeometric Functions}
\addcontentsline{toc}{subsection}{{\bf C} \ Gamma Functions and Hypergeometric Functions}
\renewcommand{\theequation}{C.\arabic{equation}}

A useful relation connecting Gamma functions which comes in handy when finding the asymptotic behaviors of various hypergeometric  functions used in the text is
\bea
\Gamma(x)\Gamma(1-x)=\frac{\pi}{\sin \pi x}
\eea
Two hypergeometric identities that we have repeatedly used in this paper are collected below:
\bea
\frac{\partial \ {}_2F_1 (a,b;c;z)}{\partial z} = \frac{ab}{c} \ {}_2F_1 (a+1,b+1;c+1;z)
{}_2
\eea
\bea {}_2F_{1}(a,b;c;z)=\frac{\Gamma(c)\Gamma(c-a-b)z^{-a}}{\Gamma(c-a)\Gamma(c-b)}{}_2F_{1}\Big(a,a-c+1;a+b-c+1;1-\frac{1}{z}\Big) +\hspace{0.6in}\\
\hspace{0.6in}\nonumber+\ \frac{\Gamma(c)\Gamma(a+b-c)}{\Gamma(a)\Gamma(b)}(1-z)^{c-a-b}z^{a-c}{}_2F_1
\Big(c-a,1-a;c-a-b+1;1-\frac{1}{z}\Big)
\eea

\subsection*{{\bf D.} \ Toric Geometry}
\addcontentsline{toc}{subsection}{{\bf D} \ Toric Geometry}
\renewcommand{\theequation}{D.\arabic{equation}}

The toric data of the $\IC^4/\IZ_4$ orbifold is useful in describing the resolution. We will be able to see purely algebraically that the singularity is replaced by a $\IC\IP^3$. %An algebraic description of the geometry can also be useful in understanding the moduli space of the dual gauge theory. 
In this appendix, we describe the $\IC^4/\IZ_4$ orbifold as a toric space. Note again that the specific complex structure we are working with is merely a crutch for deriving a smooth supergravity solution exhibiting RG flow, so one should not interpret this construction as  having a direct description of the gauge theory moduli space. A practical introduction to toric geometry can be found in \cite{greene, bouchard, Marino, Cyril}.  What we do here is an immediate generalization of \cite{suresh}. 

Before we launch into the details of the toric diagram, we first make an observation that algebraically, we can define the $\IC^4/\IZ_4$ orbifold through the algebra of the degree four monomials constructed from $w_1, w_2, w_3, w_4$ in (\ref{orbifold}). The basic idea behind any algebraic description of a space is to consider the algebra of functions over that space. In our case, there are 35 such invariant monomials that one can construct, and we can write
\bea
\IC^4/\IZ_4= \IC[P^4,Q^4,R^4,S^4,P^3Q,P^3R,P^3S,P^2Q^2,P^2QR,P^2QS,P^2R^2, \nonumber  P^2RS, P^2S^2, \\ PQ^3, PQ^2R,PQ^2S, PQR^2,PQRS,PQS^2,PR^3,PR^2S,PRS^2,PS^3,Q^3R,Q^3S,Q^2R^2  \nonumber \\
Q^2RS,Q^2S^2,QR^3,QR^2S,QRS^2,QS^3,R^3S,R^2S^2, RS^3] \equiv \IC[V_{ijkl}]\label{monom} \hspace{1.0in}
\eea
where for convenience, we have decided to use the variables $P,Q,R,S$ instead of the $w_i$'s. In the final line, we have also defined a succinct notation for the monomials for later use.

Now we turn to the toric data. We will start by writing down the vectors that define the toric diagram of the orbifold. Together with the origin, these lattice sites completely determine the space:
\bea \label{vertices}
v_1=[1,0,0,0], \ v_2=[0,1,0,0], \ v_3= [0,0,1,0], \ v_4=[-1,-1,-1,4].
\eea
To show that this is indeed the correct description of $\IC^4/\IZ_4$, we first construct the dual cones. A vector $(a,b,c,d)$ in the dual cone is defined by the condition that it has non-negative inner product with the vertices above. We want to find a set of basis vectors for the dual cones. That is, we want to find the {\em minimal} set of solutions to
\bea
a\ge 0, \ b \ge 0, \ c \ge 0, \ 4d \ge a+b+c
\eea
so that all other such vectors can be expressed as positive linear combinations of the minimal ones. To find this basis, we first notice that the first non-trivial solutions occur at $d=1$, and then we try to satisfy the inequalities in various ways. This is straightforward, and the result is
\bea
u_1=(0,0,0,1), \ u_2=(0,0,1,1), \ u_3=(0,1,0,1), \ u_4=(1,0,0,1), \ u_5=(1,1,0,1),  \nonumber \\
u_6=(1,0,1,1), \ u_7=(0,1,1,1), \ u_8=(0,0,2,1), \ u_9=(0,2,0,1), \ u_{10}=(2,0,0,1), \nonumber \\
u_{11}=(1,0,2,1), \ u_{12}=(0,1,2,1), \ u_{13}=(0,2,1,1), \ u_{14}=(1,2,0,1), \ u_{15}=(2,0,1,1), \nonumber \\
u_{16}=(2,1,0,1), \ u_{17}=(0,2,2,1), \ u_{18}=(2,0,2,1), \ u_{19}=(2,2,0,1), \ u_{20}=(2,1,1,1), \nonumber \\
u_{21}=(1,1,2,1), \ u_{22}=(1,2,1,1), \ u_{23}=(1,1,1,1), \ u_{24}=(0,0,3,1), \ u_{25}=(0,3,0,1), \nonumber \\
u_{26}=(3,0,0,1), \ u_{27}=(1,0,3,1), \ u_{28}=(0,1,3,1), \ u_{29}=(0,3,1,1), \ u_{30}=(1,3,0,1), \nonumber \\
u_{31}=(3,1,0,1), \ u_{32}=(3,0,1,1), \ u_{33}=(4,0,0,1), \ u_{34}=(0,4,0,1), \ u_{35}=(0,0,4,1). \nonumber \\
\eea
It is easy to see that the vectors in the dual cone with $d> 1$  can always be constructed with these solutions. In toric geometry, the basis of the dual cone captures the algebraic description of the original space: to each basis vector, we can associate a unique monomial (up to irrelevant overall scalings which do not affect the algebra of the monomials). In the present case, since $d=1$ for all the basis vectors, it is easy to see that with
\bea
P^{4 d-a-b-c}Q^aR^bS^c \ \ {\rm identified \ with \ the \ vector} \ \  (a,b,c,d)
\eea
the 35 basis vectors above reproduce the 35 monomials we found in (\ref{monom}). So as claimed, the toric data in (\ref{vertices}) indeed describes our orbifold algebraically.

We present the toric diagram of the orbifold in figure \ref{toric}. It is somewhat more convenient to do an $SL(4,\IZ)$ transformation on the lattice coordinates (\ref{vertices}), and we have drawn the toric diagram in the new coordinates. The new vertices are\footnote{The $SL(4,\IZ)$ matrix that accomplishes this is easy to figure out from the final and initial vertices.}
\bea \label{vertices2}
v_1=[1,0,0,1], \ v_2=[0,1,0,1], \ v_3= [0,0,1,1], \ v_4=[-1,-1,-1,1].
\eea
The advantage of the new coordinates is that now, all the vertices (except of course, the origin) have the same value for the forth coordinate.
\begin{figure}
\begin{center}
\includegraphics[%width=0.9\textwidth,
height=0.4\textheight
]{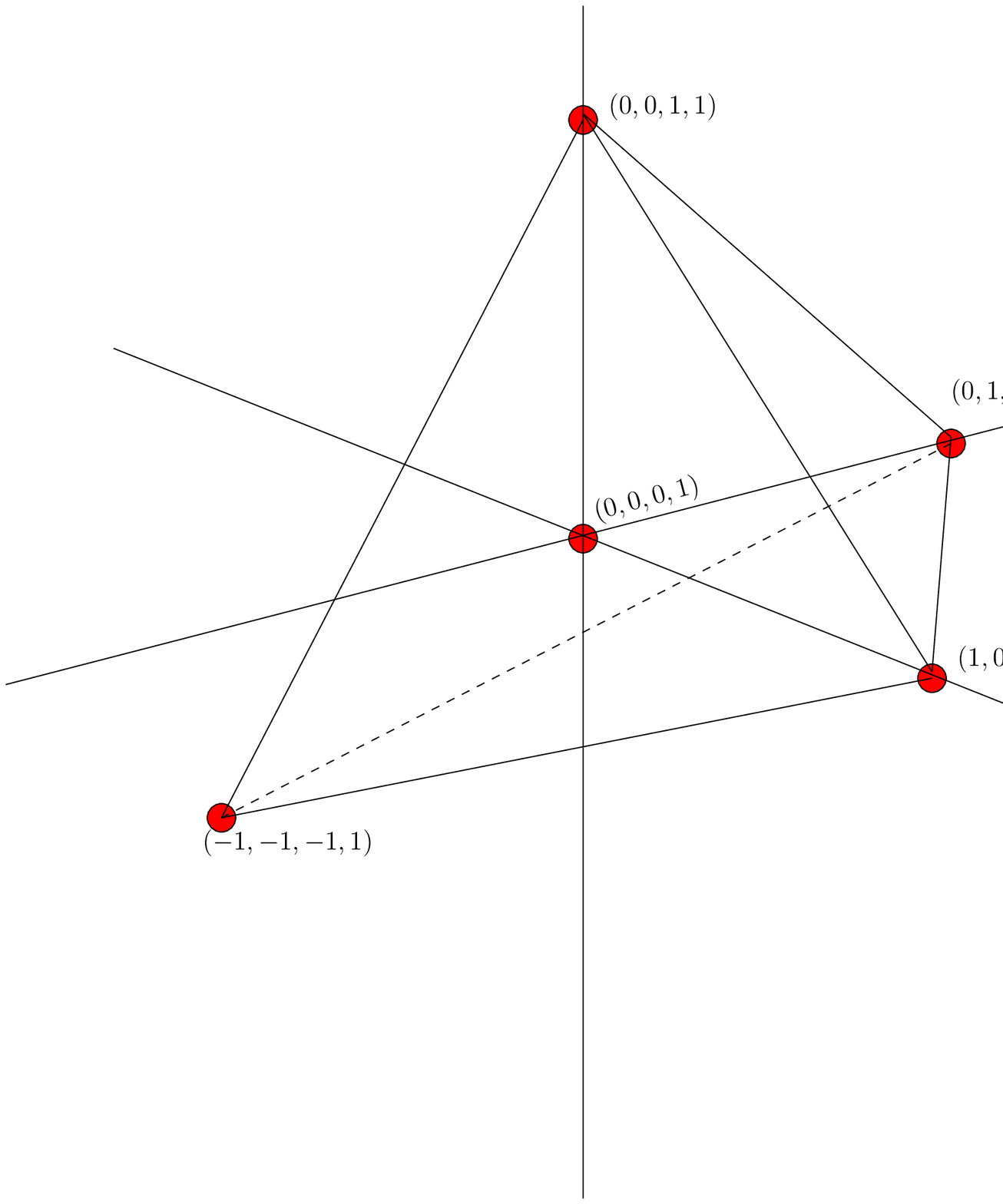}%\input{pen.pstex_t}
\caption{toric diagram of $\IC^4/\IZ_4$.}
\label{toric}
\end{center}
\end{figure}
This is possible because the original space is Calabi-Yau. In toric language, the Calabi-Yau condition translates to the condition that the vertices (other than the origin) lie on a codimension one hypersurface on the lattice. By doing the $SL(4,\IZ)$ transformation, we have made this manifest. Since the space is complex 4-dimensional, the toric diagram lives in a 4-d lattice. Ignoring the apex of the cone (namely the origin), we then have a three-dimensional (as opposed to 2-dimensional for CY 3-folds) polytope that captures the entire information about the CY space. One feature that is immediately clear from the 3-d geometry of this toric diagram is that there is a lattice point, namely $(0,0,0,1)$, that is situated in its interior. A general strategy for constructing crepant resolutions of this kind of spaces, is to add vertices corresponding to interior points and to use the new fan that is generated, as the definition of the resolved space. This means that the resolved $\IC^4/\IZ_4$ orbifold is represented, in addition to the vertices (\ref{vertices2}), by
\bea
v_5=[0,0,0,1].
\eea

With this prescription, we can go ahead and do the GLSM construction of the resolved space. The five vertices must satisfy a linear relation among them:
\bea
\sum_{i=1}^{5}Q_iv_i=0, \ \ {\rm with} \ \ Q_i=(1,1,1,1,-4)
\eea
Using these charges, we can define the resolved space in terms of complex coordinates
$\{z_1,z_2,z_3,z_4,z_5 \}$ by imposing
\bea
|z_1|^2+|z_2|^2+|z_3|^2+|z_4|^2-4|z_5|^2=\mu, \ \ \mu \ge 0, \label{Dterm}
\eea
and then modding out by the identification defined through the $U(1)$ action
\bea
(z_1,z_2,z_3,z_4,z_5) \sim (e^{i\theta}z_1,e^{i\theta}z_2,e^{i\theta}z_3,e^{i\theta}z_4,e^{-4i\theta}z_5). \label{quotient}
\eea
Note that the charges were needed to define both steps.

How do we see that the GLSM construction indeed reproduces our expectation that at the origin of the resolution, the orbifold is resolved by a $\IC\IP^3$? We first note that the $U(1)$ quotienting (\ref{quotient}) is nothing but the instruction that the basic invariant polynomials are to be constructed by multiplying $z_5$ to the degree 4 monomials constructed from $z_1, ... , z_4$. This gives rise to the monomials $z_5 V_{ijkl}$, if we identify $z_1, ... , z_4$ with $P, Q, R, S$  (see (\ref{monom})). The algebra of $z_5 V_{ijkl}$ is identical to that of $V_{ijkl}$. This means that the space is nothing but the orbifold $\IC^4/\IZ_4$, except possibly something strange happening at $z_5=0$, where the monomials collapse to zero. Since the vanishing of the monomials happened only at the orbifold point in the unresolved case, we say that the entire $z_5=0$ divisor replaces what used to be the orbifold point in the unresolved space\footnote{Depending on the value of $\mu$ the point $z_5=0$ might get resolved or not.}. Away from $z_5=0$, we have done nothing as far as the algebra is considered, so the space is unaffected. We can think of $\mu$ in (\ref{Dterm}) as the resolution parameter. To see this, note that when $\mu=0$,  $z_5=0$ forces $z_1, ... , z_4$ to be all zero, resulting in a point. But when $\mu \ne 0$, the same condition results in the usual quotient definition of $\IC\IP^3$. So we see that the resolution happens through the replacement of the orbifold point by a six-cycle.

% ==========================================================================
%
%%%%%%%%%%%%%%%%%%%%%%%%%%%%%%%%%%%%%%%%%%%%%%%%%%%%%%%%%%%%%%%%%%%%%%%%%%%%
%                      REFERENCES                            %
%%%%%%%%%%%%%%%%%%%%%%%%%%%%%%%%%%%%%%%%%%%%%%%%%%%%%%%%%%%%%%%%%%%%%%%%%%%%
\newpage
%\bibliography{metasusy}

\end{document}